\documentclass[journal]{IEEEtran}
\usepackage{silence}
\usepackage{amsmath,amssymb,amsfonts,mathtools}
\usepackage{dsfont}
\usepackage[mathscr]{euscript}
\usepackage{amsthm}
\usepackage{bbm}
\usepackage{graphicx}
\usepackage{subcaption}
\usepackage{float}
\usepackage{epstopdf}
\usepackage{color}
\usepackage{booktabs}
\usepackage{multirow}
\usepackage{array}
\usepackage[table]{xcolor}
\usepackage{tabularx}
\usepackage{diagbox}
\usepackage{algorithm}
\usepackage{algpseudocode}
\usepackage{etoolbox}
\usepackage{afterpage}
\usepackage{cite}
\usepackage{url}
\usepackage[countmax]{subfloat}
\usepackage{microtype}
\allowdisplaybreaks

\def\BibTeX{{\rm B\kern-.05em{\sc i\kern-.025em b}\kern-.08em
T\kern-.1667em\lower.7ex\hbox{E}\kern-.125emX}}

\begin{document}
\title{Secure RSMA-based Visible Light Networks under Spatial Correlation}

\author{\IEEEauthorblockN{Hung K. Hoang, 
Chuyen T. Nguyen, Thang K. Nguyen,\\ Thanh V. Pham, \textit{Senior Member, IEEE}, and 
Anh T. Pham, \textit{Senior Member, IEEE}}

\thanks{Hung K. Hoang and Chuyen T. Nguyen are with the Hanoi University of Science and Technology, Hanoi, Vietnam (e-mail: hungtpls2003@gmail.com, chuyen.nguyenthanh@hust.edu.vn).}%
\thanks{Thang K. Nguyen is an independent researcher, Hanoi, Vietnam (email: khacthang250999@gmail.com).}
\thanks{Thanh V. Pham is with Shizuoka University, Shizuoka, Japan (email: pham.van.thanh@shizuoka.ac.jp).}%
\thanks{Anh T. Pham is with The University of Aizu, Fukushima, Japan (e-mail: pham@u-aizu.ac.jp).}%
\thanks{Part of this paper has been presented at the 2024 IEEE Conference on Wireless Communications and Networking (WCNC) \cite{Thang2024}}}
\maketitle
\begin{abstract}
This paper investigates the secrecy sum rate (SSR) of rate-splitting multiple access (RSMA)-based visible light communication (VLC) systems considering internal eavesdropping, where legitimate users may intercept private data intended for others. We formulate an optimization problem to maximize the SSR of the system, which is inherently non-convex due to the complex coupling of the objective function and constraints. To this end, two different approaches based on the convex-concave procedure (CCCP) and semidefinite relaxation (SDR) are leveraged to solve the non-convex parameterized problem. A central focus of this work is the investigation of channel similarity (CS), which serves as a metric for quantifying spatial correlation, and its impact on SSR performance. To mitigate the performance degradation caused by high spatial correlation, we propose a channel similarity reduction (CSR) clustering strategy that proactively minimizes CS to restore the system's degrees of freedom (DoF). Numerical results are provided to demonstrate the performance of the two proposed algorithms under various levels of CS. More importantly, the findings reveal that our proposed CSR-clustering strategy significantly outperforms existing baselines, effectively overcoming the secrecy performance ceiling caused by high spatial correlation.
\end{abstract}  


\section{Introduction}
Visible Light Communication (VLC) has emerged as a promising solution for 6G indoor networks, exploiting the abundant unlicensed optical spectrum and existing Light-Emitting Diode (LED) infrastructure \cite{VLC_survey}. However, supporting the growing number of devices requires VLC systems to adopt efficient multiple access schemes capable of managing multi-user interference \cite{NOMA_survey, MA_VLC}. Among these, Rate-Splitting Multiple Access (RSMA) offers superior spectral efficiency and robustness compared to conventional approaches \cite{Mao2022, RSMA_LTE}. Based on rate-splitting (RS) theory \cite{Carl1978, Han1981}, RSMA mitigates interference by dividing messages into common and private parts, which are jointly precoded and superimposed. Receivers first decode the common stream via Successive Interference Cancellation (SIC) before recovering their private messages \cite{Mao2022}. By flexibly balancing interference decoding and treating it as noise, RSMA provides a unified framework that outperforms conventional Space Division Multiple Access (SDMA) and Non-Orthogonal Multiple Access (NOMA) schemes \cite{Mao2018}.

Extending its interference management capability, RSMA also offers a flexible framework for physical layer security (PLS) through its hierarchical message-splitting structure. While the common message is intended for all users, private messages must remain strictly confidential \cite{Mao2022}. However, the broadcast nature of VLC makes it inherently vulnerable to eavesdropping within the illumination footprint \cite{9070153}. In \cite{Maraqa2025}, a Reconfigurable Intelligent Surfaces (RIS)-assisted framework is proposed, where transmit beamforming and RIS phase shifts are jointly optimized to maximize the minimum secrecy rate, thereby ensuring fairness among users in the presence of external eavesdroppers. Similarly, the authors in \cite{Guo2025} develop a deep reinforcement learning approach based on a dual-sampling proximal policy optimization (DS-PPO) algorithm to maximize secrecy energy efficiency in multiple-input single-output (MISO) VLC networks, where beamforming vectors and RIS alignment matrices are adaptively optimized to counteract interception while maintaining power efficiency. Despite their effectiveness, these approaches primarily rely on reconfiguring the external propagation environment. As a result, the fundamental limitations imposed by intrinsic system geometry---such as spatial correlation and user distribution---remain largely unaddressed, potentially constraining the achievable secrecy performance in dense VLC networks.

A key manifestation of these geometric limitations is channel similarity (CS), which is defined as the cosine similarity between user channel vectors. This metric characterizes the spatial correlation between user channel vectors, with higher CS values indicating greater difficulty in distinguishing users in the spatial domain. In multi-user VLC systems, this spatial correlation leads to crosstalk that significantly degrades secrecy performance, particularly in high-CS regimes where users share overlapping optical paths. Under these conditions, the channel vectors become nearly collinear, rendering the rows of the channel matrix linearly dependent and effectively reducing its rank. Such rank deficiency reduces the available spatial degrees of freedom (DoF), fundamentally limiting the transmitter's ability to suppress inter-user interference. Since this constraint stems from the channel geometry itself rather than from power or precoder limitations, conventional beamforming and power allocation cannot recover the lost secrecy margin. This underscores the need for proactive strategies that circumvent these geometric constraints by reconfiguring the system architecture to reshape the effective channel.

To mitigate these geometric limitations by reconfiguring the system’s channel matrix, user clustering and access point (AP) association have been considered as one of the most efficient solutions. In particular, by strategically grouping users and LEDs, the network can actively manage interference beyond static physical constraints. Early works, such as the amorphous cell (A-Cell) concept in \cite{Zhang2016}, introduced user-centric clustering based on spatial location and channel gain. This user-centric paradigm is later extended with joint LED selection in \cite{Yang2022} to enhance sum-rate performance under illumination constraints. The cluster acts as a dynamic candidate pool from which only the most beneficial LEDs are activated to maximize the sum rate while adhering to strict illumination uniformity. As VLC systems evolve toward large-scale and ultra-dense deployments, clustering has shifted from local cell formation to global, multi-stage coordination. This transition is exemplified by the works of Su \textit{et al}., where a stable matching framework was introduced in \cite{Su2024}. Unlike conventional greedy approaches, it models user equipments (UEs) and APs as two sets of players with preference lists, ensuring stable associations. This framework was further extended in \cite{Su2025} using spectral clustering. The network is modeled as a weighted graph, where vertices represent UEs (or clusters) and edge weights capture mutual interference. From this graph, the Laplacian matrix is constructed to encode the network’s connectivity structure, and its eigenvectors are used to project the nodes into a low-dimensional space where strongly interfering nodes lie close together. Clustering in this space groups highly correlated users into the same sub-network, thereby confining interference and reducing inter-sub-network interference prior to resource allocation. Similar advanced coordination is explored in \cite{Ihsan2025,Tuan2025}, which employ hybrid TDMA-NOMA clustering to exploit channel disparities while orthogonalizing users in time. These approaches are complemented by coalitional grouping \cite{Papani2019}, interference-aware clustering \cite{Liu2020a, Liu2020b}, and energy-efficient amorphous designs for video streaming \cite{Li2017, Obeed2019}.
However, by prioritizing channel gain or load balancing, these strategies often perform poorly when users are in close proximity. In such cases, overlapping optical footprints lead to high CS, resulting in strongly correlated channels regardless of AP association. This severely limits the ability to securely separate users' signals, necessitating CS-aware designs to ensure reliable and secure transmission in highly correlated environments.

To address these limitations, this paper develops robust precoding strategies to maximize the secrecy sum rate (SSR) in multi-user RSMA-based VLC systems under a worst-case secrecy model, where each user is treated as a potential eavesdropper. The resulting non-convex design problem is tackled via two efficient suboptimal precoding schemes that achieve reliable local optima. Moreover, to overcome the impact of the CS on the designs, we further propose a Channel Similarity Reduction (CSR) clustering framework that proactively reconfigures user–LED associations prior to precoding. By jointly maximizing channel gain and minimizing inter-user correlation through a multi-objective meta-heuristic, the proposed approach restores spatial separability and enables robust secure transmission. To the best of our knowledge, this is the first work to explicitly investigate secrecy-aware clustering in RSMA-based VLC systems. The main contributions of this paper are summarized as follows.
\vspace{-0.03cm}
\begin{itemize}
    \item We formulate the SSR maximization problem for RSMA-based MU-MISO VLC systems under a worst-case internal eavesdropping model, where legitimate users may collude to intercept private streams intended for others. The resulting optimization problem is highly non-convex due to the coupled secrecy-rat expressions, RS power allocation constraints, and VLC amplitude limitations.
    \item To address the multivariate non-convexity of the SSR maximization problem, we develop two suboptimal design approaches based on the Convex-Concave Procedure (CCCP) and Semidefinite Relaxation (SDR). A dedicated Zero-Forcing-Maximum Ratio Transmission (ZF-MRT) initialization strategy is introduced to significantly reduce computational overhead and ensure rapid convergence.
    \item More importantly, we demonstrate that high CS fundamentally reduces the effective rank of the VLC channel matrix, thereby collapsing the spatial DoF required for secure precoding. In such regimes, conventional beamforming and power allocation strategies become insufficient to suppress information leakage, resulting in a secrecy-performance ceiling even at high transmit power.
    \item To overcome this geometry-induced limitation, we propose a novel CSR clustering framework that proactively restores spatial separability prior to precoding. Specifically, the proposed method jointly partitions users and LEDs using a constrained multi-objective Non-dominated Sorting Genetic Algorithm II (NSGA-II) that simultaneously maximizes channel strength while minimizing intra-cell channel correlation. By reshaping the effective channel geometry, the proposed framework enlarges the feasible secure-precoding region and restores the spatial DoF necessary for robust RSMA transmission.
    \item Extensive simulations reveal that the proposed CSR framework significantly outperforms conventional user-centric clustering strategies, particularly in dense VLC deployments with strong spatial correlation. The results further reveal that RSMA can partially alleviate DoF deficiency through interference decoding, while the proposed CSR design is essential for overcoming the secrecy saturation caused by geometry-induced channel correlation.
\end{itemize}

The remainder of the paper is organized as follows. Section \ref{sec:model} presents the system model, including the VLC channel representation, the RSMA transmission structure, the secrecy rate formulation, and the concept of CS and RS power allocation. Section \ref{sec:non_clustered} develops two precoding strategies for maximizing the SSR and analyzes their computational complexity. Section \ref{sec:cluster} introduces the proposed CS-reduction framework based on joint user clustering and LED assignment. Representative simulation results are provided in Section \ref{sec:simu}, and finally, Section \ref{sec:conclusion} concludes the paper.

\begin{figure*}[ht]
\centerline{\includegraphics[scale = 0.32]{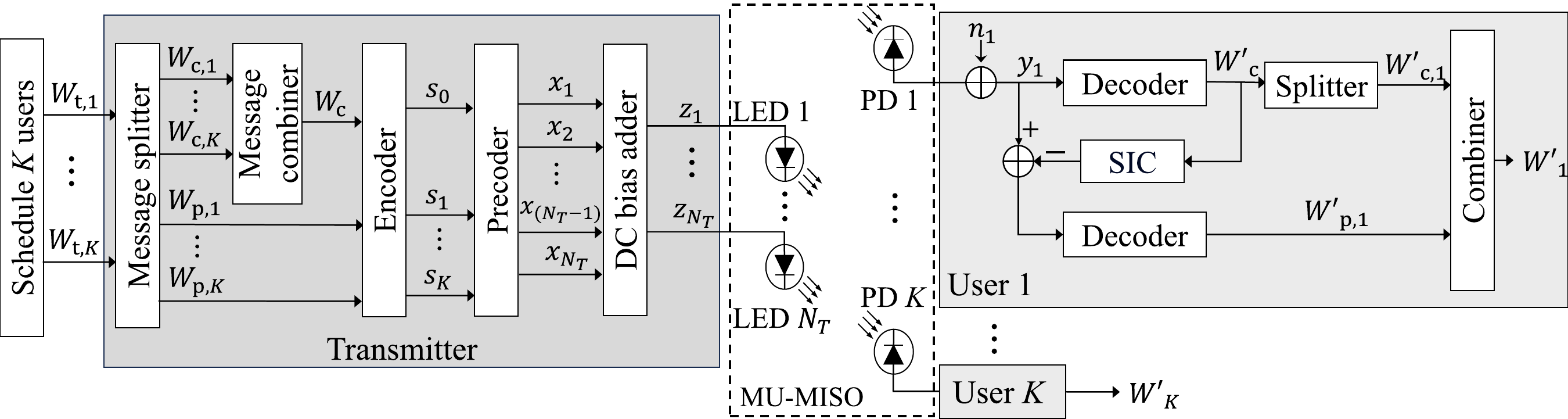}}
\caption{Schematic diagram of MU-MISO RSMA-based VLC system.}
\label{system_model}
\vspace{-\baselineskip}
\end{figure*}

\textit{Notation:} Italic, bold lowercase, and bold uppercase letters denote scalars, column vectors, and matrices, respectively. The space of $M \times N$ real-valued matrices is denoted by $\mathbb{R}^{M \times N}$. For a matrix $\mathbf{X}$, $[\mathbf{X}]_{i,j}$ and $[\mathbf{X}]_{i,:}$ represent its $(i,j)$-th entry and $i$-th row, respectively. The operators $(\cdot)^{\text{T}}$, $\text{Tr}(\cdot)$, and $\mathrm{vec}(\cdot)$ denote the transpose, trace, and vectorization, respectively. For a vector $\mathbf{x}$, $\|\mathbf{x}\|$ and $\|\mathbf{x}\|_1$ denote the Euclidean norm and the $\ell_1$-norm, respectively. Additionally, sets are represented by calligraphic fonts (e.g., $\mathcal{S}$), where $|\mathcal{S}|$ denotes the cardinality of set $\mathcal{S}$. Finally, $\mathbbm{1}_{\{\cdot\}}$ is the indicator function.

\section{System Model}
\label{sec:model}
\subsection{VLC System Description}
As illustrated in Fig.~\ref{system_model}, we consider a downlink multi-user multiple-input single-output (MU-MISO) VLC system employing RSMA. The transmitter is equipped with $N_T$ LED luminaries serving $K$ independent users, each equipped with a photodiode (PD). The LED array is responsible for delivering both illumination and secure data transmission. Each user acts not only as a legitimate user to decode its desired message but also as a potential eavesdropper to wiretap messages transmitted to other users. The VLC signal propagates from the transmitter to the receiver via both line-of-sight (LoS) and non-line-of-sight (NLoS) paths, which include reflections. However, as the LoS component typically contributes more than 95\% of the total received optical power \cite{Komine2004}, only the LoS propagation path is considered in this work.

The VLC channel gain $h_{n,k}$ between the $n$-th LED luminary and the $k$-th user is characterized by the standard Lambertian model, and can be expressed as \cite{Komine2004} 
\vspace{-0.1cm}
\begin{align} 
\label{eqn:chann_coeff}
    h_{n,k} = \frac{A_r}{d_{n,k}^2}L(\phi_{n,k})T_s(\psi_{n,k})g(\psi_{n,k})\cos(\psi_{n,k}), 
\end{align}
where $A_r$ is the active area of the PD at the receiver, $d_{n,k}$ is the Euclidean distance between the $n$-th LED and the $k$-th user, $\phi$ and $\psi$ represent the angle of irradiance and angle of incidence, respectively. Furthermore, $L(\phi) = \frac{l+1}{2\pi}\cos^l(\phi)$ is the Lambertian emission intensity, where $l = -\frac{\ln(2)}{\ln(\Theta_{0.5})}$ denotes the Lambertian index, with $\Theta_{0.5}$ being the LED's semi-angle at half illuminance. Besides, $T_s(\psi_{n,k})$ is the gain of optical filter, and $g(\psi_{n,k})$ denotes the gain of the optical concentrator, which can be calculated as $g(\psi_{n,k}) = \mathbbm{1}_{\{0 \leq \psi_{n,k} \leq \Psi\}} \kappa^2/\sin^2(\Psi)$ \cite{Komine2004}, where $\kappa$ is the refractive index of the concentrator and $\Psi$ denotes the optical field-of-view (FOV) of the PD.

\vspace{-0.2cm}
\subsection{MU-MISO RSMA Transmission Model}
In the considered RSMA transmission model, the message intended for the $k$-th user, denoted by $W_{\mathrm{t},k}$, is divided into two parts: a common part $W_{\mathrm{c},k}$, which is jointly decoded by all users, and a private part $W_{\mathrm{p},k}$, decoded only by its intended user. The common parts of all users’ messages are combined into $W_{\mathrm{c}}$ and encoded into a common stream $s_0$ using a shared codebook, while each private part $W_{\mathrm{p},k}$ is independently encoded into a private stream $s_k$. It is assumed that the data symbols $s_k$ are zero-mean with variance $\sigma_s^2$, and is normalized to the range of $[-1,1]$. These $(K + 1)$ streams $\mathbf{s} = [s_0 \quad s_1 \quad s_2 \quad \ldots \quad s_K]^{\text{T}}$ are linearly precoded to form the transmitted electrical signal denoted by $\mathbf{x}\in \mathbb{R}^{N_T \times 1}$ and is expressed as $\mathbf{x} = \mathbf{P}\mathbf{s} = \sum_{k=0}^K \mathbf{p}_k s_k$, where $\mathbf{P} = [\mathbf{p}_0 \quad \mathbf{p}_1 \quad \ldots \quad \mathbf{p}_K] \in \mathbb{R}^{N_T \times (K+1)}$ denotes the precoding matrix.

To ensure the non-negativity of the optical intensity required for proper LED operation, a DC bias vector $\mathbf{I}^{\mathrm{DC}} = [I_1^{\mathrm{DC}} \quad I_2^{\mathrm{DC}} \quad \ldots \quad I_{N_T}^{\mathrm{DC}}]^{\mathrm{T}}$ is added to $\mathbf{x}$. The resulting LED drive current vector $\mathbf{z}\in \mathbb{R}^{N_T \times 1}$ is given by $\mathbf{z} = \mathbf{x}\, + \,\mathbf{I}^{\mathrm{DC}},$ which must satisfy the illumination constraint $0 \leq z_n \leq I_{\max},\; (n = 1, 2, \ldots, N_T),$ where $I_{\max}$ denotes the maximum permissible drive current ensuring linear LED operation. Given that the transmitted symbols are normalized to $|s_k| \le 1$, the {signal} amplitude constraint can be equivalently expressed as
\begin{equation} \label{eqn:amplitude_constraint}
    \lVert [\mathbf{P}]_{n,:} \rVert_1 \le \min\big(I_n^{\mathrm{DC}}, I_{\max} - I_n^{\mathrm{DC}}\big),
\end{equation}
where $\lVert\cdot\rVert_1$ denotes the $\ell_1$-norm and $[\mathbf{P}]_{n,:}$ is the $n$-th row of $\mathbf{P}$. The emitted optical power of the $n$-th LED luminary is then given by $P_n^s = \eta z_n = \eta (x_n + I_n^{\mathrm{DC}}),$ where $\eta$ denotes the electro–optical conversion efficiency.

Let $\mathbf{H} \in \mathbb{R}^{K \times N_T}$ denote the VLC channel matrix, whose $k$-th row $\mathbf{h}_k = [h_{k,1} \quad h_{k,2} \quad \ldots \quad h_{k,N_T}]$ represents the LoS channel vector between all $N_T$ LEDs and the $k$-th user. The total optical signal received at the $k$-th PD can be expressed as
\begin{align} \label{eqn:rx_signal_1}
    y_k &= \gamma \, \mathbf{h}_k [P_1^s \quad P_2^s \quad \ldots \quad P_{N_T}^s]^{\text{T}} + n_k
\end{align}
\begin{equation} \label{eqn:rx_signal_2}
        \notag = \gamma \eta \left( \mathbf{h}_k \mathbf{p}_0 s_0 + \mathbf{h}_k \mathbf{p}_k s_k + \mathbf{h}_k\hspace{-0.2cm} \sum_{i=1,\, i \ne k}^{K} \hspace{-0.2cm}\mathbf{p}_i s_i+ \mathbf{h}_k \mathbf{I}^{\mathrm{DC}}
        \right) +\, n_k,
\end{equation}
where $\gamma$ denotes the PD responsivity and $n_k$ represents additive white Gaussian noise (AWGN) with zero mean and variance $\sigma_k^2$, which is given by \cite{zeng2009}. The DC component $\gamma \eta \mathbf{h}_k \mathbf{I}^{\mathrm{DC}}$ is removed at the receiver via AC coupling, yielding the received AC signal as \vspace{-0.4cm}
\begin{equation}
\label{eqn:rx_AC_signal}
    \vspace{-0.1cm} \overline{y}_k = \mathbf{h}_k \mathbf{p}_0 s_0 + \mathbf{h}_k \mathbf{p}_k s_k + \mathbf{h}_k \sum_{i=1,\, i \ne k}^{K} \mathbf{p}_i s_i + \overline{n}_k,
\end{equation}
where $\overline{n}_k = \frac{n_k}{\gamma \eta}$ denotes the normalized electrical noise.

\subsection{Rate Analysis}
At the $k$-th user, the common stream $s_0$ is first decoded while treating all private streams as interference. Hence, a lower bound on the achievable common rate can be obtained via the mutual information between $s_0$ and the received signal $\overline{y}_k$, expressed as
\begin{align} \label{eqn:R_ck_mutual}
    R_{\mathrm{c},k}(\mathbf{P}) &\ge \mathbb{I}(s_0; \overline{y}_k) = h(\overline{y}_k) - h(\overline{y}_k|s_0) 
    \\&= h\Big(\mathbf{h}_k \sum_{i=0}^K \mathbf{p}_i s_i + \overline{n}_k\Big) - h\Big(\mathbf{h}_k \sum_{i=1}^K \mathbf{p}_i s_i + \overline{n}_k\Big), \nonumber
\end{align}
where $\mathbb{I}(\cdot;\cdot)$ denotes mutual information, while $h(\cdot)$ and $h(\cdot|\cdot)$ are the differential and conditional entropies, respectively. To obtain a tractable expression, we apply the Gaussian bounding method, which approximates both the received signal and the interference-plus-noise term as Gaussian variables with the same variance. The resulting closed-form lower bound on the common rate is given by
\begin{equation} \label{eqn:R_ck}
    R_{\mathrm{c},k}(\mathbf{P}) = \tfrac{1}{2}\log_2\!\bigg(\frac{1 + \sum_{i=0}^K a_k (\mathbf{h}_k \mathbf{p}_i)^2}{1 + \sum_{i=1}^K b_k (\mathbf{h}_k \mathbf{p}_i)^2}\bigg),
\end{equation}
where $a_k = \frac{2^{2h(s_k)}}{2\pi e\,\overline{\sigma}_k^2},\, b_k = \frac{\sigma_s^2}{\overline{\sigma}_k^2}, \, \overline{\sigma}_k^2 = \frac{\sigma_k^2}{(\gamma\eta)^2}.$

It is noted that to ensure successful decoding of the common stream by all users, the transmission rate of the common stream denoted by $R_c$ must not exceed the minimum achievable common rate across users, i.e., $ R_c \le \min_{k\in\{1,\ldots,K\}} R_{\mathrm{c},k}(\mathbf{P}).$ After the common stream $s_0$ is decoded and subtracted by successive interference cancellation (SIC), the residual received signal at the $k$-th user is given by

\vspace{-0.3cm}
\begin{equation} 
    \label{eqn:rx_SIC_signal}
    \overline{y}_{\mathrm{p},k} = \mathbf{h}_k \mathbf{p}_k s_k + \mathbf{h}_k \mathbf{\hat{P}}_k \mathbf{\hat{s}}_k + \overline{n}_k,
\end{equation}
where $\mathbf{\hat{s}}_k$ and $\mathbf{\hat{P}}_k$ are, respectively, obtained from $\mathbf{s}$ and $\mathbf{P}$ by removing the entries corresponding to $s_0$ and $s_k$. The received signal vector of the remaining $K-1$ users after removing the common stream is written as
\begin{equation} \label{eqn:rx_remaining_users}
    \mathbf{\hat{y}}_{\mathrm{p},k} = \mathbf{\hat{H}}_k \mathbf{p}_k s_k + \mathbf{\hat{H}}_k \mathbf{\hat{P}}_k \mathbf{\hat{s}}_k + \mathbf{\hat{n}}_k,
\end{equation}
where $\mathbf{\hat{H}}_k \in \mathbb{R}^{(K-1)\times N_T}$ is obtained by removing the $k$-th row from $\mathbf{H}$, and $\mathbf{\hat{n}}_k = [\overline{n}_1 \quad \cdots \quad \overline{n}_{k-1} \quad \overline{n}_{k+1} \quad \cdots \quad \overline{n}_K]^{\mathrm{T}}$. To guarantee data confidentiality, we adopt a conservative physical-layer security model in which the other $K-1$ users may collude to eavesdrop on $s_k$. Following the same bounding methodology as above, a closed-form lower bound on the secrecy rate of the $k$-th user can be expressed as \cite{EE_Son_2021}
\begin{align} \label{eqn:R_sk}
    R_{s,k}(\mathbf{P}) &\ge \mathbb{I}(s_k; \overline{y}_{\mathrm{p},k}) - \mathbb{I}(s_k; \mathbf{\hat{y}}_{\mathrm{p},k} \mid \mathbf{\hat{s}}_k) - \mathbb{I}(s_k; \mathbf{\hat{s}}_k) \notag
    \\&= \tfrac{1}{2}\log_2\!\bigg(\frac{1 + \sum_{i=1}^K a_k (\mathbf{h}_k \mathbf{p}_i)^2}{1 + \sum_{i=1,\,i\ne k}^K b_k (\mathbf{h}_k \mathbf{p}_i)^2} \bigg) \notag
    \\&- \tfrac{1}{2}\log_2\!\bigg( 1 + \sum_{i=1,\,i\ne k}^K b_i (\mathbf{h}_i \mathbf{p}_k)^2 \bigg),
\end{align}
where the first term represents the achievable rate at the legitimate user after common-message cancellation, and the second term upper bounds the information leakage to the set of colluding eavesdroppers. The last term $\mathbb{I}(s_k; \mathbf{\hat{s}}_k)$ is zero under the assumption that data streams are independently encoded.

Combine (\ref{eqn:R_ck}) and (\ref{eqn:R_sk}), we obtain the secrecy sum rate (SSR) of the system as
\vspace{-0.2cm}
\begin{align} 
\label{eq:SSR}
    \Phi(\mathbf{P}) = R_c \, &+ \, \tfrac{1}{2}\log_2\!\bigg(\frac{1 + \sum_{i=1}^K a_k (\mathbf{h}_k \mathbf{p}_i)^2}{1 + \sum_{i=1,\,i\ne k}^K b_k (\mathbf{h}_k \mathbf{p}_i)^2} \bigg) \notag
    \\&- \tfrac{1}{2}\log_2\!\bigg( 1 + \sum_{i=1,\,i\ne k}^K b_i (\mathbf{h}_i \mathbf{p}_k)^2\bigg).
\end{align}
\subsection{Channel Similarity and RS Power Allocation}

\subsubsection{Channel Similarity}
In a MU-MISO VLC system, the channel matrix $\mathbf{H}$ depends on the spatial geometry and orientation of the LED array and user PDs. Each user $k$ has a channel gain vector $\mathbf{h}_k$ determined by its line-of-sight distance, irradiance angle, and incidence angle. When users are spatially well separated, their channel vectors tend to be orthogonal, leading to a well-conditioned and full-rank $\mathbf{H}$ that enables efficient precoding. Conversely, closely spaced or symmetrically located users yield highly correlated channels, reducing spatial DoF and potentially degrading both secrecy and interference suppression. To quantify this inter-user channel correlation, we employ the CS metric based on the normalized cosine similarity between user channel vectors. Specifically, for given $K$ users with channel matrix $\mathbf{H} = [\mathbf{h}_1, \ldots, \mathbf{h}_K]$, the average CS is defined as
\begin{equation} \label{eq:avg_cs}
\mathrm{CS}(\mathbf{H}) =\frac{2}{K(K-1)} \sum_{i<k}\frac{\mathbf{h}_i\cdot \mathbf{h}_k} {\lVert \mathbf{h}_i \rVert \lVert \mathbf{h}_k \rVert},
\end{equation}
where $\lVert \cdot \rVert$ denotes the Euclidean norm.
The CS value lies in $[0,1]$, where lower values indicate better spatial separability and reduced inter-user interference.
\vspace{0.3em}
\subsubsection{RS Power Allocation}
In RSMA systems, the transmit power is allocated between the common and private streams to manage the trade-off between interference suppression and efficient resource utilization. The relative distribution of power across these two components strongly affects system performance, including achievable rate region and overall energy efficiency~\cite{Mao2022,EE_RSMA_2018}. A higher power share to the common stream enhances decoding robustness at all users but may limit the flexibility of individual link adaptation, whereas allocating more power to the private streams can improve user-specific throughput at the cost of increased inter-user interference. Let $P_{\mathrm{c}}$ and $P_{\mathrm{p}}$ denote the transmit power allocated to the common and private streams, respectively. The RS power-allocation ratio is defined as
\begin{equation} \label{eqn:power-allocation}
    \rho = \frac{P_{\mathrm{p}}}{P_{\mathrm{c}}} = \frac{\sum_{k=1}^{K}\lVert\mathbf{p}_k\rVert^2}{\lVert\mathbf{p}_0\rVert^2},
\end{equation}
where $\mathbf{p}_k$ and $\mathbf{p}_0$ are the precoding vectors of the $k$-th private stream and the common stream, respectively.  

\section{SSR Maximization}
\label{sec:non_clustered}
We now turn our attention to designing the optimal precoding that maximizes the SSR in (\ref{eq:SSR}). In particular, we consider the following $\mathbb{P}\mathbf{1}$

\begin{table}[ht]
\begin{small}
\begin{subequations}
\label{original_problem}
    \begin{flalign} 
        \mathbb{P} \mathbf{1}:~ 
        \underset{\substack{\mathbf{P}, R_\text{c}, \mathbf{x}, \mathbf{y}}}{\text{\normalsize maximize}} 
        \hspace{3mm} & \Phi(\mathbf{P}) \label{objectiveP1}
        && 
    \end{flalign}      
    \vspace{-1.5\baselineskip}
    \hspace{-0.65 cm}
    \begin{minipage}{0.2\textwidth}
        \vspace{-0.5cm}
        \begin{align}
        & \nonumber\text{\normalsize subject to} \\& R_{s,k}(\textbf{P}) \geq \lambda_k, \label{P1:constraint_Rs} 
        \\& R_{c,k}(\textbf{P}) \geq R_c, \label{P1:constraint_Rc}
        \\& \xi \text{Tr}(\mathbf{P} \mathbf{P}^{\text{T}}) \leq P_\text{t},\label{P1:constraint_Pt}
        \end{align}
    \end{minipage}
    \hfill
    \begin{minipage}{0.3\textwidth}
        \begin{align}    &\sum\textstyle_{k=1}^K\lVert\mathbf{p}_k\rVert^2 = \rho \lVert\mathbf{p}_0\rVert^2,       \label{P1:constraint_PA} 
                \\&     \left\lVert\left[\mathbf{P}\right]_{n, :}\right\rVert_1 \hspace{-0.1cm} \leq \min\left(I_n^{\text{DC}},I_{\text{max}}-I_n^{\text{DC}}\right), \label{P1:constraint_amplitude}
        \end{align}
    \end{minipage}\bigskip
\end{subequations}
\end{small}
\end{table}

\noindent where \eqref{P1:constraint_Rs} guarantees that the achievable secrecy rate of the $k$-th user remains above the predefined minimum threshold $\lambda_k$, \eqref{P1:constraint_Rc} follows from the definition $R_{\text{c}} = \min_{k}R_{\text{c},k}(\mathbf{P})$, \eqref{P1:constraint_Pt} constrains the total transmit power allocated to both common and private messages, \eqref{P1:constraint_PA} enforces the power-allocation relationship specified in \eqref{eqn:power-allocation}, and  \eqref{P1:constraint_amplitude} is the signal amplitude constraint. At \eqref{P1:constraint_Pt}, $\xi = r\sigma_s^2$ with $r$ being the equivalent AC resistance. It is recognized that (\ref{original_problem}) is a non-convex problem because of the non-convexity of (\ref{objectiveP1}), (\ref{P1:constraint_Rs}), (\ref{P1:constraint_Rc}), and (\ref{P1:constraint_PA}), which makes it difficult to solve. To cope with this problem, in the following, two solution approaches are proposed. The first approach employs a series of variable transformations and first-order Taylor approximations to obtain a locally convex surrogate of problem (\ref{original_problem}), which is then efficiently solved using the concave–convex procedure (CCCP) in an iterative manner \cite{yuille2003}. The second approach leverages the semidefinite relaxation (SDR) technique to reduce the number of auxiliary variables and the degree of approximation required. In this case, the resulting problem is expressed in a difference-of-convex (d.c.) programming form, which can likewise be efficiently handled via the CCCP framework \cite{Horst1999}.

\subsection{CCCP approach}
It is noticed that (\ref{objectiveP1}) can be transformed to be concave by introducing the following slack variables: $x_{1,k} \triangleq \frac{1}{2} \log_2 ( 1 + y_{1,k} )$ with $y_{1,k} \triangleq \sum_{i=0}^{K} a_k ( \mathbf{h}_k \mathbf{p}_i )^2$, $x_{2,k} \triangleq \frac{1}{2} \log_2 ( 1 + y_{2,k} )$ with $y_{2,k} \triangleq \sum_{i=1}^{K} b_k ( \mathbf{h}_k \mathbf{p}_i )^2$, $x_{3,k} \triangleq \frac{1}{2} \log_2 ( 1 + y_{3,k} )$ with $y_{3,k} \triangleq \sum_{i=1}^{K} a_k ( \mathbf{h}_k \mathbf{p}_i )^2$, $x_{4,k} \triangleq \frac{1}{2} \log_2 ( 1 + y_{4,k} )$ with $y_{4,k} \triangleq \sum_{i=1, i\neq k}^{K} b_k ( \mathbf{h}_k \mathbf{p}_i )^2$, and $x_{5,k} \triangleq \frac{1}{2} \log_2 ( 1 + y_{5,k} )$ with $y_{5,k} \triangleq \sum_{i=1, i \neq k}^{K} b_i ( \mathbf{h}_i \mathbf{p}_k )^2$. Accordingly, $\mathbb{P}\mathbf{1}$ can be equivalently reformulated to $\mathbb{P}\mathbf{2}$ as shown below. 

\vspace{-0.4cm}
\begin{table}[ht]
\begin{small}
\begin{subequations}
    \label{CCP_problem}
    \begin{flalign} 
        \mathbb{P} \mathbf{2}:~ 
        \underset{\substack{\mathbf{P}, R_\text{c}, \mathbf{x}, \mathbf{y}}}{\text{maximize}} 
        \hspace{3mm} & R_{\text{c}} + \sum_{k=1}^K 
        (x_{3,k} - x_{4,k} - x_{5,k}), 
        && \label{objectiveCCP}
    \end{flalign}      
    \vspace{-1.5\baselineskip}
    \hspace{-0.5 cm}
    \begin{minipage}{0.25\textwidth}
        \vspace{-1 cm}
        \begin{align}
        & \nonumber\text{subject to}  \\
        & R_\text{c} ~\leq x_{1,k} - x_{2,k},
        \label{P2:constraint_CCPb}\\
        & x_{1,k} \leq \frac{1}{2}\log_2\left(1+y_{1,k}\right),            \label{P2:constraint_CCPc} \\
        & x_{2,k} \geq \frac{1}{2}\log_2\left(1+y_{2,k}\right),            \label{P2:constraint_CCPd} \\
        & x_{3,k} \leq \frac{1}{2}\log_2\left(1+y_{3,k}\right),           \label{P2:constraint_CCPe} \\
        & x_{4,k} \geq \frac{1}{2}\log_2\left(1+y_{4,k}\right),            \label{P2:constraint_CCPf} \\
        & x_{5,k} \geq \frac{1}{2}\log_2\left(1+y_{5,k}\right),            \label{P2:constraint_CCPg} \\
        & x_{3,k} - x_{4,k} - x_{5,k} \geq \lambda_k, \label{P2:constraint_CCPh}
        \\& \eqref{P1:constraint_PA}, \eqref{P1:constraint_amplitude}, 
        \eqref{P1:constraint_Pt}, \nonumber
        \end{align}
    \end{minipage}
    \hfill
    \begin{minipage}{0.24\textwidth}
        \vspace{-0.2cm}
        \begin{align}
        &  y_{1,k} \leq \sum_{i=0}^K a_k\left(\mathbf{h}_k^\text{T} 
        \mathbf{p}_i\right)^2,     
        \label{P2:constraint_CCPi} \\
        & y_{2,k} \geq \sum_{i=1}^K b_k\left(\mathbf{h}_k^\text{T} \mathbf{w}_i\right)^2,     
        \label{P2:constraint_CCPj}\\
        & y_{3,k} \leq \sum_{i=1}^K a_k\left(\mathbf{h}_k^\text{T} \mathbf{p}_i\right)^2,     
         \label{P2:constraint_CCPk}\\
        & y_{4,k} \geq \hspace{-0.23 cm} \sum_{i=1,i\neq k}^K \hspace{-0.23 cm} b_k\left(\mathbf{h}_k^\text{T} \mathbf{p}_i\right)^2,     
        \label{P2:constraint_CCPl}\\
        & y_{5,k} \geq \hspace{-0.23 cm} \sum_{i=1,i\neq k}^K \hspace{-0.23 cm} b_i\left(\mathbf{h}_k^\text{T} \mathbf{p}_k\right)^2.     \label{P2:constraint_CCPm}
        \end{align}
    \end{minipage}\bigskip
\end{subequations}
\end{small}
\vspace{-0.2cm}
\end{table}

It is noted in $\mathbb{P}\mathbf{2}$ that the equalities defined above are relaxed into their corresponding inequality constraints (\ref{P2:constraint_CCPc})–(\ref{P2:constraint_CCPg}) and (\ref{P2:constraint_CCPi})–(\ref{P2:constraint_CCPm}), respectively. This relaxation does not affect the optimality of the problem. Specifically, since the objective function in (\ref{objectiveCCP}) is monotonically increasing with respect to $x_{1,k}$, its optimal value must occur when $x_{1,k}$ attains its upper bound, meaning that (\ref{P2:constraint_CCPc}) and (\ref{P2:constraint_CCPi}) hold with equality at optimality. The same reasoning applies to the remaining variables, for which the corresponding inequality constraints are also active at the optimum. It is observed that all constraints are convex except for \eqref{P2:constraint_CCPd}, \eqref{P2:constraint_CCPf}, \eqref{P2:constraint_CCPg}, \eqref{P2:constraint_CCPi}, \eqref{P2:constraint_CCPk}, and \eqref{P1:constraint_PA}. To handle these non-convex terms, first-order Taylor expansions are applied to construct locally tight convex surrogates at the $(m-1)$-th iteration. Specifically, $\mathbf{p}_k^{(m-1)}$, $y_{2,k}^{(m-1)}$, $y_{4,k}^{(m-1)}$, and $y_{5,k}^{(m-1)}$ represent the values of $\mathbf{p}_k$, $y_{2,k}$, $y_{4,k}$, and $y_{5,k}$ obtained from the previous iteration, respectively. With these linearizations, problem $\mathbb{P}\mathbf{2}$ is reformulated as the convex optimization problem as follows, and can be efficiently solved using standard convex optimization tools such as \texttt{CVX}.\cite{GrantBoydCVX2014}

\vspace{-0.4cm}
\begin{small}
\begin{subequations}
\label{CCP_problem_final}
\begin{align}
    &\mathbb{P} \mathbf{3}:
    \nonumber\underset{\substack{\mathbf{P}, R_\text{c}, \mathbf{x}, \mathbf{y}}}{\text{maximize}} \quad (\ref{objectiveCCP})  
    \\&\text{subject to} \nonumber 
    \\&  \qquad y_{1, k} \leq \sum_{i=0}^K a_k \mathcal{L}^{(1)}(\mathbf{p}_i, \mathbf{p}_i^{(m-1)}), \label{P3:constraint_CCPa} 
    \\& \qquad x_{j, k} \geq \mathcal{L}^{(2)}(y_{j,k},y_{j,k}^{(m-1)}), \, j \in \{2,4,5\}, \label{P3:constraint_CCPb_x3} 
    \\& \qquad y_{3, k} \leq \sum_{i=1}^K a_k \mathcal{L}^{(1)}(\mathbf{p}_i, \mathbf{p}_i^{(m-1)}), \label{P3:constraint_CCPc} 
    \\& \qquad \sum_{k=1}^K \mathcal{L}^{(1)}(\mathbf{p}_k, \mathbf{p}_k^{(m-1)}) = \rho \mathcal{L}^{(1)}(\mathbf{p}_0, \mathbf{p}_0^{(m-1)}), \label{P3:constraint_CCPd} 
    \\&  \qquad
    \eqref{P2:constraint_CCPb},    \eqref{P2:constraint_CCPc},
      \eqref{P2:constraint_CCPe}, \eqref{P2:constraint_CCPh},
      \eqref{P2:constraint_CCPj}, \eqref{P2:constraint_CCPl},
      \eqref{P2:constraint_CCPm}, \eqref{P1:constraint_amplitude},
      \eqref{P1:constraint_Pt}, \nonumber
\end{align}
\end{subequations}
\end{small}

\noindent where $\mathcal{L}^{(1)}(\mathbf{a}, \mathbf{a}^{(m-1)}) \triangleq \|\mathbf{a}^{(m-1)}\|^2 + 2(\mathbf{a}^{(m-1)})^{\text{T}}(\mathbf{a} - \mathbf{a}^{(m-1)})$ and $\mathcal{L}^{(2)}(y_{j,k},y_{j,k}^{(m-1)}) = \frac{1}{2}\log_2(1+y_{j,k}^{(m-1)}) + \frac{y_{j,k}-y_{j,k}^{(m-1)}}{2\ln(2)(1+y_{j,k}^{(m-1)})}.$ The solution to $\mathbb{P}\mathbf{2}$ is then obtained iteratively by applying the CCCP algorithm, which successively solves $\mathbb{P}\mathbf{3}$ until convergence. The overall CCCP-based procedure for solving $\mathbb{P}\mathbf{2}$ is summarized in \textbf{Algorithm~\ref{alg:CCCP}}.

\vspace{-0.15cm}
\begin{algorithm}[ht]
\small
\caption{CCCP algorithm for solving $\mathbb{P}\mathbf{2}$.}
\label{alg:CCCP}
\begin{algorithmic}[1]
\State Choose the maximum number of iterations $N_{\text{max},1}$ and the error tolerance $\epsilon_1 > 0$.
\State Initialize feasible starting points $\mathbf{P}^{(0)}$, $y^{(0)}_{2,k}$, $y^{(0)}_{4,k}$, and $y^{(0)}_{5,k}$.
\State Set $m \gets 0$.
\While{convergence == \textbf{False} \text{and} $m \le N_{\text{max},1}$}
    \State Use $\mathbf{P}^{(m-1)}$, $y^{(m-1)}_{2,k}$, $y^{(m-1)}_{4,k}$, $y^{(m-1)}_{5,k}$ obtained from the 
    \Statex \quad \, previous iteration to formulate problem $\mathbb{P}\mathbf{4}$.
    \State Solve $\mathbb{P}\mathbf{4}$ to get $\mathbf{P}^{(m)}$, $y^{(m)}_{2,k}$, $y^{(m)}_{4,k}$, and $y^{(m)}_{5,k}$.
    \If{
        $\dfrac{\|\mathbf{P}^{(m)} - \mathbf{P}^{(m-1)}\|}{\|\mathbf{P}^{(m)}\|} \le \epsilon_{1}$
        \textbf{and}
        $\dfrac{|y^{(m)}_{i,k} - y^{(m-1)}_{i,k}|}{y^{(m)}_{i,k}} \le \epsilon_{1}$, for \\ \quad \, $i \in \{2,4,5\}$
    }
        \State convergence $\gets$ \textbf{True}
        \State $\mathbf{P}^* \gets \mathbf{P}^{(m)}$
        \State $y_{2,k}^* \gets y_{2,k}^{(m)}$, \; $y_{4,k}^* \gets y_{4,k}^{(m)}$, \; $y_{5,k}^* \gets y_{5,k}^{(m)}$
    \Else
        \State convergence $\gets$ \textbf{False}
    \EndIf
    \State $m \gets m + 1$
\EndWhile
\State \textbf{Return} the optimal value $\mathbf{P}^* \gets \mathbf{P}^{(m)}$.
\end{algorithmic}
\end{algorithm}

\vspace{-0.4cm}
\subsection{Combined CCCP-SDR approach}
This approach leverages SDR to reformulate the non-convex problem $\mathbb{P}\mathbf{1}$. In particular, let $\mathbf{P}_k \triangleq \mathbf{h}_k^{\text{T}} \mathbf{h}_k$, $\mathbf{Q}_0 \triangleq \mathbf{w}_0 \mathbf{w}_0^{\text{T}}$, and $\mathbf{Q}_k \triangleq \mathbf{w}_k \mathbf{w}_k^{\text{T}}$ (for $k \in \{1, \dots, K\}$), by which these new variables must satisfy $\mathbf{Q}_k \succeq 0$ and the non-convex constraint $\text{rank}(\mathbf{Q}_k)=1$. Then, using the property $(\mathbf{h}_k \mathbf{w}_i)^2 = \text{Tr}(\mathbf{P}_k \mathbf{Q}_i)$, the common rate in \eqref{eqn:R_ck} can be rewritten as 
\begin{equation}
    \label{eq:R_ck_Q}
    R_{c,k}(\mathbf{Q}) = f_{c,k}(\mathbf{Q}) - g_{c,k}(\mathbf{Q}),
\end{equation}
where $\mathbf{Q} \triangleq \{\mathbf{Q}_0, \dots, \mathbf{Q}_K\}$, and $f_{c,k}(\mathbf{Q})$ and $g_{c,k}(\mathbf{Q})$ are both concave functions as follows
\begin{align}
    \label{fg_ck}
    f_{c,k}(\mathbf{Q}) &= \frac{1}{2}\log_2\left(1 + \sum_{i=0}^{K}a_k \text{Tr}(\mathbf{P}_k\mathbf{Q}_i)\right), \\
    g_{c,k}(\mathbf{Q}) &= \frac{1}{2}\log_2\left(1+\sum_{i=1}^{K}b_k\text{Tr}(\mathbf{P}_k\mathbf{Q}_i) \right).
\end{align}

As $R_{c,k}$ is now a d.c. function, we again employ the CCCP algorithm. Specifically, at the $m$-th iteration, the concave term $g_{c,k}(\mathbf{Q})$ is approximated by its first-order Taylor approximation around the solution from the previous iteration $\mathbf{Q}^{(m-1)}$ as
\begin{align}
    \label{g_ck_approx}
    g_{c,k}(\mathbf{Q}) \approx & \ g_{c,k}(\mathbf{Q}^{(m-1)}) \nonumber \\
    & + \, \sum_{i=1}^K\left\langle \nabla_{g_{c,k}}\left(\mathbf{Q}_i^{(m-1)}\right),\mathbf{Q}_i-\mathbf{Q}_i^{(m-1)}\right\rangle_\text{F},
\end{align}

\noindent where $\langle\cdot,\cdot\rangle_\text{F}$ denotes the Frobenius inner product. The gradient $\nabla_{g_{c,k}}\left(\mathbf{Q}_i^{(m-1)}\right)$ evaluated at point $\mathbf{Q}_i^{(m-1)}$ is calculated as
\begin{equation}
    \label{grad_gck}
\nabla_{g_{c,k}}\left(\mathbf{Q}_i^{(m-1)}\right) = \mathbbm{1}_{\{i > 0\}} z_{c,k} b_k \mathbf{P}_k,
\end{equation}
where $z_{c,k}$ is given by
\begin{equation}
    \label{z_ck}
    z_{c,k}^{(m-1)} = \frac{1}{2\ln(2)\left(1+\sum_{j=1}^{K}b_k\text{Tr}\left(\mathbf{P}_k\mathbf{Q}_j^{(m-1)}\right) \right)}.
\end{equation}

On the other hand, the secrecy rate \eqref{eqn:R_sk} of the $k$-th user is similarly reformulated as a d.c. function as
\begin{equation}
    R_{s,k}(\mathbf{Q}) = f_{s,k}(\mathbf{Q}) - g_{s,k}(\mathbf{Q}),   
\end{equation}
where
\begin{align}
    & f_{s,k}(\mathbf{Q}) = \frac{1}{2} \log_2 \left(1 + \sum_{i=1}^{K} a_k\text{Tr}(\mathbf{P}_k\mathbf{Q}_i)\right),    \label{f_sk}
    \\& g_{s,k}(\mathbf{Q}) =  \frac{1}{2}\log_2 \left(1 + \sum_{i=1, i \neq k}^{K} b_k \text{Tr}(\mathbf{P}_k\mathbf{Q}_i)\right) \nonumber 
    \\& \qquad \quad \;+ \frac{1}{2} \log_2 \left( 1 + \sum_{i=1, i \neq k}^{K} b_i \text{Tr}(\mathbf{P}_i\mathbf{Q}_k)\right).    \label{g_sk}
\end{align}
At the $m$-th iteration, the concave term $g_{s,k}(\mathbf{Q})$ is also approximated using its first-order Taylor expansion:
\begin{align}
    g_{s,k}(\mathbf{Q}) \approx & \ g_{s,k}(\mathbf{Q}^{(m-1)}) \nonumber \\
    & + \sum_{i=1}^K \left\langle \nabla_{g_{s,k}}\left(\mathbf{Q}_i^{(m-1)}\right), \mathbf{Q}_i - \mathbf{Q}_i^{(m-1)} \right\rangle_\text{F},
\end{align}
where the gradient $\nabla_{g_{s,k}}\left(\mathbf{Q}_i^{(m-1)}\right)$ is calculated as:
\begin{align}
    \nabla_{g_{s,k}}\left(\mathbf{Q}_i^{(m-1)}\right) & =  \mathbbm{1}_{\{i \neq 0\}} \big[ \mathbbm{1}_{\{i \neq k\}} (z_{s,k}^1)b_k\mathbf{P}_k \notag 
    \\& + \mathbbm{1}_{\{i = k\}} (z_{s,k}^2)\sum_{j=1, j\neq k}^K b_j\mathbf{P}_j \big],
\end{align}
where $z_{s,k}^1$ and $z_{s,k}^2$ are given by
\begin{align}
    & z_{s,k}^1 = \frac{1}{2\ln(2) \left(1+\sum_{j=1,j \neq k}^{K}b_k\text{Tr}\left(\mathbf{P}_k\mathbf{Q}_j^{(m-1)}\right)\right)}, \\
    & z_{s,k}^2 = \frac{1}{2\ln(2) \left(1+\sum_{j=1,j \neq k}^{K}b_j\text{Tr}\left(\mathbf{P}_j\mathbf{Q}_k^{(m-1)}\right)\right)}.
\end{align}

By linearizing the concave components $g_{c,k}(\mathbf{Q})$ and $g_{s,k}(\mathbf{Q})$, the objective function, the constraints \eqref{P1:constraint_Rs} and \eqref{P1:constraint_Rc} become convex. Furthermore, the constraints \eqref{P1:constraint_PA} and \eqref{P1:constraint_Pt} can be re-written, respectively, as $\sum_{k=1}^K \text{Tr}(\mathbf{Q}_k) = \rho\text{Tr}(\mathbf{Q}_0),$ and $\xi \sum_{i=0}^K \text{Tr}(\mathbf{Q}_i) \leq P_t,$ which are linear (and thus convex) with respect to $\mathbf{Q}_i$. The remaining challenge is the constraint \eqref{P1:constraint_amplitude}, which is an $\ell_1$-norm constraint. Here, since it is difficult to represent $\sum_{k=1}^K |w_{n,k}|$ directly in terms of $\mathbf{Q}_k$, we adopt the sequential approximation proposed in \cite{EE_Son_2021}. In particular, the constraint is replaced by a stricter, convex surrogate that is updated at each CCCP iteration $m$ using the solution from the previous one $\mathbf{w}_k^{(m-1)}$. The resulting constraint is given by \cite{EE_Son_2021}

\vspace{-0.3cm}
\begin{small}
\begin{equation}
\label{eqn:constraint_amplitude_rewrite}
\sum_{k=0}^K \frac{\text{Tr}(\mathbf{E}_n\mathbf{Q}_k)}{\sqrt{\text{Tr}\left(\mathbf{E}_n\mathbf{Q}_k^{(t-1)}\right)}} \leq \frac{\left( \min\left(I_n^{\text{DC}},I_{max}- I_n^{\text{DC}}\right)^2 \right)}{\sum_{k=0}^K \sqrt{\text{Tr}\left(\mathbf{E}_n\mathbf{Q}_k^{(t-1)}\right)}},
\end{equation}
\end{small}
where $\mathbf{E}_n \in \mathbb{R}^{N_T \times N_T}$ is an all-zero matrix with a $1$ at the $(n,n)$-th entry. The problem can then be reformulated with respect to $\mathbf{Q}$ in $\mathbb{P}_4$ as follow

\vspace{-0.45cm}
\begin{small}
\begin{subequations}
\label{SDR_problem}
\begin{alignat}{2}
    & \hspace{-0.75cm} \mathbb{P}\mathbf{4}:  \underset{\mathbf{Q},R_c}{\text{maximize}} \quad  R_c + \sum_{k=1}^K \left[ f_{s,k}(\mathbf{Q}) - \mathcal{L}^{(3)}_{s,k}(\mathbf{Q}, \mathbf{Q}^{(t-1)}) \right], \label{objectiveSDR} 
    \\& \hspace{-0.75cm} \text{subject to} \nonumber
    \\&  \hspace{-0.75cm} f_{s,k}(\mathbf{Q}) - \mathcal{L}^{(3)}_{s,k}(\mathbf{Q}, \mathbf{Q}^{(t-1)}) \ge \lambda_k, \quad \forall k, \label{P4_constraint_SDRa} 
    \\& \hspace{-0.75cm} f_{c,k}(\mathbf{Q}) - \mathcal{L}^{(3)}_{c,k}(\mathbf{Q}, \mathbf{Q}^{(t-1)}) \ge R_c, \quad \forall k, \label{P4_constraint_SDRb}
\end{alignat}
\hspace{-0.46cm}
\vspace{0.1cm}
\begin{minipage}{0.25\textwidth}
    \vspace{-\baselineskip}
    \begin{align}
        & \sum_{k=1}^K \text{Tr}(\mathbf{Q}_k) = \rho \text{Tr}(\mathbf{Q}_0), \label{eq:P4_42d} 
        \\&\mathbf{Q}_k \succeq 0, \quad \forall k , \label{eq:P4_42g}
    \end{align}
\end{minipage}
\hfill
\begin{minipage}{0.2\textwidth}
    \vspace{-\baselineskip}
    \begin{align}
        & \xi \sum_{k=0}^{K} \text{Tr}(\mathbf{Q}_k) \leq P_t, \label{eq:P4_42e}
        \\&(\ref{eqn:constraint_amplitude_rewrite}), \nonumber
    \end{align}
\end{minipage}
\end{subequations}
\end{small}

\noindent where $\mathcal{L}^{(3)}_{m,k}(\mathbf{Q}, \mathbf{Q}^{(t-1)})  \triangleq g_{m,k}(\mathbf{Q}^{(t-1)}) + \sum_{i=1}^K \langle \nabla_{g_{m,k}}$ $(\mathbf{Q}_i^{(t-1)}), \mathbf{Q}_i - \mathbf{Q}_i^{(t-1)} \rangle_F, \; \forall m \in \{s,c\}.$ The algorithm to solve $\mathbb{P}_4$ is given in \textbf{Algorithm \ref{alg:SDR}.}

\begin{algorithm}[ht]
\small
\caption{CCCP-SDR algorithm for solving $\mathbb{P}\mathbf{4}$.}
\label{alg:SDR}
\begin{algorithmic}[1]
\State Choose the maximum number of iterations $N_{\text{max},2}$ and the error tolerance $\epsilon_2 > 0$.
\State Choose an initial point $\mathbf{P}^{(0)}$ being feasible to the problem (\ref{original_problem}) and calculate corresponding matrices $\mathbf{Q}_0^{(0)},\mathbf{Q}_1^{(0)},...,\mathbf{Q}_K^{(0)}$
\State Set $m \gets 0$.
\While{convergence == \textbf{False} \text{and} $m \le N_{\text{max},2}$}
    \State Use $\mathbf{Q}_0^{(m-1)},\mathbf{Q}_1^{(m-1)},...,\mathbf{Q}_K^{(m-1)}$ that obtained from the 
    \Statex \quad \; previous iteration to formulate the problem (\ref{SDR_problem}).
    \State Solve $\mathbb{P}\mathbf{4}$ to get $\mathbf{Q}_0^{(m)*},\mathbf{Q}_1^{(m)*},...,\mathbf{Q}_K^{(m)*}$.
    \State Use (\ref{eq:rank-one approx}) to get $\mathbf{Q}_0^{(m)},\mathbf{Q}_1^{(m)},...,\mathbf{Q}_K^{(m)}$ as rank-one approxi-
    \Statex \quad \, mation to $\mathbf{Q}_0^{(m)*},\mathbf{Q}_1^{(m)*},...,\mathbf{Q}_K^{(m)*}$, respectively.
    \State Retrieve $\mathbf{P}^{(m)}$ from $\mathbf{Q}_i^{(m)}, i \in \overline{0,K}$.
    \If{
        $\dfrac{\|\mathbf{P}^{(m)} - \mathbf{P}^{(m-1)}\|}{\|\mathbf{P}^{(m)}\|} \le \epsilon_{2}$
    }
        \State convergence $\gets$ \textbf{True}
        \State $\mathbf{P}^* \gets \mathbf{P}^{(m)}$
    \Else
        \State convergence $\gets$ \textbf{False}
    \EndIf
    \State $m \gets m + 1$
\EndWhile
\State \textbf{Return} the optimal value $\mathbf{P}^*$.
\end{algorithmic}
\end{algorithm}

It is worth mentioning that the optimal solution of (\ref{SDR_problem}) obtained at each iteration may not inherently satisfy the rank-one constraints. Therefore, in the proposed \textbf{Algorithm~\ref{alg:SDR}}, the matrix solution from the $m$-th iteration, denoted by $\mathbf{Q}_k^{(m)*}$, is further projected onto a rank-one space. The resulting approximation, denoted by $\mathbf{Q}_k^{(m)}$, is computed as follows.
\begin{equation}
\label{eq:rank-one approx}
    \mathbf{Q}_k^{(m)} = \Lambda^{(m)}_{\text{max},k}\mathbf{q}^{(m)}_{\text{max},k}\left[\mathbf{q}^{(m)}_{\text{max},k} \right]^{\text{T}},
\end{equation}
where $\Lambda^{(m)}_{\text{max},k}$ is the maximum eigenvalue of $\mathbf{Q}_k^{(m)*}$ and $\mathbf{q}^{(m)}_{\text{max},k}$ is the eigenvector associated with $\Lambda^{(m)}_{\text{max},k}$.

\subsection{Complexity Analysis}
In each iteration, the Interior Point Algorithm (IPA) \cite {Nesterov1994} is assumed to be adopted to solve the corresponding convex sub-problems, i.e., (\ref{CCP_problem_final}) and (\ref{SDR_problem}). In this case, the computational complexity is characterized by the total number of Newton steps (denoted as $N_s^{\text{total}}$) required to reach the convergence, which can be expressed as $N_s^{\text{total}} = N_{\text{iter}} \times N_s,$ where $N_{\text{iter}}$ denotes the total number of iterations and $N_s$ represents the number of Newton steps executed in each iteration. According to \cite{Nesterov1994} and \cite{Arfaoui2018}, the worst-case number of Newton steps for solving a nonlinear convex problem scales approximately with the square root of the number of scalar variables, i.e., $N_s \sim \sqrt{\text{number of scalar variables}}.$

For problem (\ref{CCP_problem_final}), the optimization variables include $\mathbf{P}$, $R_c$, $\mathbf{x}$, and $\mathbf{y}$, where $\mathbf{x}$ and $\mathbf{y}$ each contains $(5\times K)$ scalar entries, and $\mathbf{P} \in \mathbb{R}^{N_T \times (K+1)}$. Hence, the overall Newton steps can be approximated as
\begin{equation}
    N_s^{\text{\textbf{Algorithm 1}}} \sim \sqrt{K(N_T + 10) + N_T+1}.
\end{equation}

In contrast, problem (\ref{SDR_problem}) involves $(K+1)$ matrix variables ${\mathbf{Q}_i}$, each of size $\mathbb{R}^{N_T\times N_T}$ and a scalar variable $R_c$. Thus, the corresponding Newton steps scale as
\begin{equation}
    N_s^{\text{\textbf{Algorithm 2}}} \sim \sqrt{N_T^2 (K + 1) + 1}.
\end{equation}

It can be observed that the computational load of \textbf{Algorithm~2} grows much faster than that of \textbf{Algorithm~1}, particularly as the number of transmit LEDs increases. This is because \textbf{Algorithm~2} involves $N_T \times N_T $ matrix variables, whereas \textbf{Algorithm~1} only deals with vector-valued parameters.

\section{Joint User–LED Partitioning for Channel Similarity Reduction (CSR)}
\label{sec:cluster}
In this section, to address multi-user scenarios where high CS may degrade system performance, we develop a joint partitioning framework that determines the user–LED association topology (i.e., the cell layout) \cite{Verma2021}. Each cell is assigned a dedicated frequency sub-band to ensure negligible inter-cell interference \cite{Chen2015, Elsayed2023}. By ensuring sufficient linear independence among co-channel users, this structural optimization restores the spatial DoF, thereby establishing the necessary conditions for the subsequent precoding stage to yield a non-trivial SSR. Although this configuration simplifies interference management, it raises concerns regarding spectral efficiency and system complexity, which we leave for future investigation. Moreover, while the proposed framework can, in principle, be extended to an arbitrary number of cells $C$, for clarity and tractability, we restrict our analysis to the two-cell case ($C = 2$). Our primary goal is to identify network configurations that balance maximizing received optical signal strength and minimizing intra-cell correlation. Once such configurations are determined, the optimal precoding strategy from the previous section can be applied.

\subsection{Problem Formulation}
We consider the problem of simultaneously partitioning the set of users $\mathcal{U} =\{1,2,...,K\}$ and the set of LEDs $\mathcal{L}=\{1,2,..,.N_T\}$ into two disjoint cells, denoted by $\mathcal{C}_1$ and $\mathcal{C}_2$. Each cell $\mathcal{C}_c$ is associated with a subset of users $\mathcal{U}_c \subseteq \mathcal{U}$ and a subset of LEDs $\mathcal{L}_c \subseteq \mathcal{L}$ such that $\mathcal{U}_1\, \cup \, \mathcal{U}_2 = \mathcal{U}$ and $\mathcal{U}_1\, \cap \, \mathcal{U}_2 = \emptyset$; $\mathcal{L}_1\, \cup \, \mathcal{L}_2 = \mathcal{L}$ and $\mathcal{L}_1\, \cap \, \mathcal{L}_2 = \emptyset$. Define a joint partition of users and LEDs as $\mathcal{S} = \{(\mathcal{U}_1,\mathcal{L}_1),(\mathcal{U}_2,\mathcal{L}_2) \}$ and all sets are disjoint. For each cell, let $\mathbf{H}_c$ be the channel sub-matrix associated with the user set $\mathcal{U}_c$ and LED set $\mathcal{L}_c$. The proposed clustering strategy aims to improve the signal quality with inter-cell fairness, while mitigating spatial correlation to support secure transmission. To formalize these design objectives, we introduce two objective functions, as follows

\vspace{-0.15cm}
\begin{minipage}{0.2\textwidth}
    \begin{equation}
        \hspace{-0.57cm}
        f_1(\mathcal{S}) = \prod_{c=1}^2 \|\mathrm{vec}(\mathbf{H}_c)\|_1, \nonumber
    \end{equation}
\end{minipage}
\hfill
\begin{minipage}{0.2\textwidth}
    \begin{equation}
        f_2(\mathcal{S}) = \sum_{c=1}^2 \left[ \mathrm{CS}(\mathbf{H}_c) \right]^2. \nonumber
    \end{equation}
\end{minipage}
\noindent 
The function $f_1(\mathcal{S})$ promotes strong channel conditions by maximizing the aggregate channel gain. The product form further enforces inter-cell fairness, as it penalizes imbalanced configurations in which one cell dominates in power. In contrast, $f_2(\mathcal{S})$ captures the level of spatial correlation by summing the squared correlation measures across channels. Its minimization suppresses correlated pairings, preserving spatial DoF crucial for secure precoding.

These two objectives are, however, inherently antagonistic. Maximizing $f_1$ encourages users and LEDs to be placed in close spatial proximity to harvest higher optical power, whereas minimizing $f_2$ requires greater spatial separation to preserve channel orthogonality. As a result, no single solution can simultaneously optimize both objectives; instead, a set of optimal trade-offs, i.e., the Pareto front \cite{Konak2006}, exists. In this case, scalarization techniques (e.g., weighted-sum methods) are often employed. However, these approaches are fundamentally limited, as they fail to identify unsupported efficient solutions in the non-convex regions (i.e., concavities) of the Pareto front \cite{Jaszkiewicz2002}, due to the combinatorial nature of user-LED clustering. Furthermore, determining appropriate weights for incommensurable objectives is highly subjective and often leads to biased solutions that fail to reflect the true trade-off characteristics \cite{Gunantara2018}. To overcome these theoretical limitations and capture the complete Pareto front without bias, we formulate the problem as a Constrained Multi-Objective Combinatorial Optimization (CMOCO) problem as follows \cite{Hao2024, Verma2021}

\begin{small}
\begin{subequations} \label{prob:joint_MOP} 
\begin{align}
    \mathbb{P}\mathbf{5}: \underset{\mathcal{S}}{\text{minimize}} \quad &  \{-f_1(\mathcal{S}),\;  f_2(\mathcal{S}) \}
    \\ \text{subject to} \quad &\max_{c \in {1,2}} \big(\mathrm{CS}(\mathbf{H}_c)\big) \leq \tau, \label{const:CS} 
    \\ &   |\mathcal{U}_c| \geq 2, \quad \forall c, \label{const:min_user} 
    \\ &   |\mathcal{L}_c| \geq |\mathcal{U}_c|, \quad \forall c, \label{const:mimo} 
    \\ &   (|\mathcal{U}_1| - |\mathcal{U}_2|) (|\mathcal{L}_1| - |\mathcal{L}_2|) \geq 0. \label{const:prop} 
\end{align} 
\vspace{-\baselineskip}
\end{subequations}
\end{small}

In this formulation, (\ref{const:CS}) ensures the channel correlation in any cell does not exceed a predefined threshold $\tau$, (\ref{const:min_user}) requires each cell to serve at least 2 users to justify the use of RSMA techniques. Moreover, (\ref{const:mimo}) ensures the number of transmit LEDs is at least equal to the number of users in each cell, providing sufficient DoF for precoding and (\ref{const:prop}) enforces a fair resource distribution by requiring that the cell with more users is assigned a greater (or equal) number of LEDs. 

The optimization problem $\mathbb{P}_5$ involves a discrete, high-dimensional search space with non-linear dependencies, rendering it NP-hard. Furthermore, the handling of strict structural constraints classifies $\mathbb{P}_5$ as a challenging Constrained Multi-Objective Problem (CMOP) \cite{Hao2024}. According to recent comprehensive surveys \cite{Verma2021, Hao2024}, Constraint-Handling Techniques integrated within evolutionary frameworks like Non-dominated Sorting Genetic Algorithm II (NSGA-II) have proven to be one of the most robust approaches for such problems. Specifically, NSGA-II is favored for its capability to maintain population diversity in the presence of complex infeasible regions \cite{Konak2006, Deb2002}. Therefore, we adopt a Modified NSGA-II framework to identify the Pareto-optimal user–LED partitions.

\subsection{Key Components of Modified NSGA-II Framework}
{The direct application of standard continuous-domain optimization algorithms is ineffective in this setting, as their genetic operators inherently produce non-integer solutions that violate the discrete constraints in (\ref{const:CS})–(\ref{const:prop})}. To address this limitation, we introduce a modified NSGA-II framework with four key adaptations: (i) an integer-based encoding scheme to represent user–cell associations; (ii) the Constrained-Domination Principle to ensure solution feasibility; (iii) customized genetic operators to maintain valid network structures during the evolution i.e., selection, crossover, and mutation; and (iv) crowding distance to maintain the population diversity. The detailed implementation of these adaptations is described below.
\subsubsection{Chromosome Representation}
A candidate solution is encoded as a joint integer-valued chromosome $\mathbf{x}$ of length $K + N_T$. The chromosome is structured as the concatenation of the user partition and the LED partition vectors, written as
\begin{equation}
    \mathbf{x} = [\underbrace{u_1,...,u_K}_{\text{User}}\,|\,\underbrace{l_1,...,l_{N_T}}_{\text{LED}}]^{\text{T}},
\end{equation}
where $u_k, l_n \in \{1,2\}$ implies the assignment of the $k$-th user, $n$-th LED to their respective cells. Let $\mathcal{S}(\mathbf{x})$ represent the physical partition decoded from $\mathbf{x}$ and thus, the objective values associated with $\mathbf{x}$ are evaluated as $f_m(\mathcal{S}(\mathbf{x}))$. For simplicity, we will henceforth denote these values simply as $f_m(\mathbf{x})$.

\subsubsection{Constraint Handling}
The Constrained-Domination Principle (CDP) \cite{Deb2002} is incorporated into the NSGA-II framework to enforce solution feasibility through a feasibility-based ranking strategy. Unlike conventional NSGA-II implementations that rank individuals solely according to Pareto dominance in the objective space, CDP explicitly accounts for constraint satisfaction during the non-dominated sorting process. Specifically, any feasible solution is always ranked higher than an infeasible one, regardless of objective values. When comparing infeasible solutions, preference is given to those with smaller overall constraint violations. To implement this mechanism, we define a scalar constraint-violation measure, denoted by $V(\mathbf{x})$, which aggregates the violations of all unmet constraints and is expressed as
\begin{align}
V(\mathbf{x}) &= \;\max(0, \max_c(\text{CS}_c) - \tau) + \mathbbm{1}_{\{\min_{c}|\mathcal{U}_c| < 2\}} \\&
+\; \mathbbm{1}_{\{|\mathcal{L}_c| < |\mathcal{U}_c|\}}+ \mathbbm{1}_{\left\{\left(|\mathcal{U}_1| - |\mathcal{U}_2|\right)\left(|\mathcal{L}_1| - |\mathcal{L}_2|\right) < 0\right\}}, \nonumber
\end{align}
A solution $\mathbf{x}_i$ is said to constrained-dominate $\mathbf{x}_j$ if {any of the following conditions is met}
\begin{itemize}
    \item $\mathbf{x}_i$ is feasible $ (\text{\small $V$}(\mathbf{x}) = 0)$ and $\mathbf{x}_j$ is infeasible $(V(\mathbf{x}) > 0)$,
    \item both are infeasible, but $\mathbf{x}_i$ has a smaller violation $(V(\mathbf{x}_i)  < V(\mathbf{x}_j))$, 
    \item both are feasible, and $\mathbf{x}_i$ Pareto-dominates $\mathbf{x}_j$ in terms of objectives $\{f_1,f_2\}$, meaning that it improves at least one objective without degrading the others.
\end{itemize}

\subsubsection{Genetic Operators (GO)} The evolution of the population is driven by three main operators:

\textbf{Selection:} Binary tournament selection is performed based on the CDP. In each tournament, two individuals are randomly selected, and the one that constrained-dominates the other is chosen as a parent. If neither individual constrained-dominates the other, the selection is decided by the crowding distance (which will be described later), with preference given to the individual exhibiting a larger crowding distance to preserve population diversity \cite{Deb2002}.

\textbf{Joint Crossover with Repair: }Offspring solutions, denoted by $\mathbf{x}_{\text{child}}$ are generated using uniform crossover, where each gene (corresponding to a user or an LED) is inherited from either parent $\mathbf{x}_{p_1}$ and $\mathbf{x}_{p_2}$ with equal probability. Due to the stochastic nature of the crossover operation, the resulting offspring may violate structural constraints. To address this issue, and following established practices for modified NSGA-II implementations \cite{Verma2021}, we incorporate a two-stage repair mechanism to restore feasibility.
\begin{itemize}
    \item Stage 1 (User Rebalancing): If the offspring violates the minimum user requirement in \eqref{const:min_user}, users are iteratively migrated from the majority cell to the minority cell until the cardinality constraint is satisfied.    
    \item Stage 2 (LED Alignment): First, for any cell $c$ violating \eqref{const:mimo}, LEDs are reassigned from the alternative cell until the requirement is met. Subsequently, the proportionality constraint in \eqref{const:prop} is enforced by iteratively transferring LEDs from the cell with a surplus (relative to its associated user load) to the cell with a deficit.
\end{itemize}

\textbf{Mutation:} With a small probability $p_m$ (for exploration without destroying a good solution), a randomly selected gene in either the user or LED segment is reassigned to the alternative cell. This operation enables the discovery of alternative clusterings, the correction of early misassignment, and thus, prevents premature convergence \cite{Konak2006}.

\subsubsection{Crowding Distance}
To preserve diversity along the Pareto front, the Crowding Distance (CD) metric is employed. This metric estimates the local density of solutions surrounding a given individual in the objective space, where a larger CD value indicates that the solution lies in a sparsely populated region and is therefore preferred to avoid solution clustering. The CD is computed independently for each objective. Specifically, for each objective $m$, the population is sorted in ascending order of $f_m$, and the boundary solutions are assigned an infinite CD. For the remaining intermediate solutions, the crowding distance $d_i$ is accumulated across objectives as \cite{Deb2002}
\begin{equation}
    d_i = \sum_{m=1}^2 \frac{f_m(\mathbf{x}_{i+1}) - f_m(\mathbf{x}_{i-1})}{f_m^{\max} - f_m^{\min}},
\end{equation}
where $f_m(\mathbf{x}_{i\pm1})$ denotes the objective values of the immediate neighbors of solution $i$ in the sorted list, and $f_m^{\max/\min}$ represents the normalization bounds for objective $m$. The crowding distance $d_i$ provides an estimate of the local density of solutions surrounding $x_i$ in the objective space by measuring the normalized distance between its nearest neighbors, thereby promoting a well-distributed approximation of the Pareto front.

\subsection{Algorithm Execution and Selection Strategy}
\subsubsection{Algorithm Procedure} Let $\mathcal{P}_t$ denote the population of candidate solutions at generation $t$. In addition, let $N_{\text{pop}}$ and $N_{\text{gen}}$ be the population size and the number of generations, respectively. The execution flow of the proposed modified NSGA-II is formally summarized in \textbf{Algorithm \ref{alg:NSGAII}}.
The algorithm starts by initializing a feasible population $\mathcal{P}_0$, where each chromosome represents a valid joint user–LED assignment satisfying constraints \eqref{const:CS}–\eqref{const:prop}. At each generation, offspring are generated through the proposed genetic operators, evaluated, and combined with the parent population. The resulting population is then ranked using non-dominated sorting with the CDP, followed by crowding-distance-based environmental selection to form the next generation. This process is repeated until the termination criterion is met.
\begin{algorithm}[ht]
\small
\caption{Proposed NSGA-II for User-LED Clustering}
\label{alg:NSGAII}
\begin{algorithmic}[1]
\renewcommand{\algorithmicrequire}{\textbf{Input:}}
\renewcommand{\algorithmicensure}{\textbf{Output:}}
\Require Population size $N_{\text{pop}}$, max generations $N_{\text{gen}}$.
\Ensure Feasible Pareto front $\mathcal{F}_1$.
\State $\mathcal{P}_0 \gets \emptyset$
\While{$|\mathcal{P}_0| < N_{\text{pop}}$}
    \State Generate random chromosome $\mathbf{x}$;
    \If{$\mathbf{x}$ satisfies constraints \eqref{const:CS}-\eqref{const:prop}}
        \State $\mathcal{P}_0 \gets \mathcal{P}_0 \cup \{\mathbf{x}\}$;
    \EndIf
\EndWhile
\State Evaluate $\{f_1, f_2\}$ and violation $V(\mathbf{x})$ for all $\mathbf{x} \in \mathcal{P}_0$;
\For{$t = 1$ to $N_{\text{gen}}$}
    \State $\mathcal{Q}_t \gets \emptyset$;
    \State \textbf{Apply} Genetic Operators on $\mathcal{P}_{t-1}$ to fill $\mathcal{Q}_t$;
    \State Evaluate $\{f_1, f_2\}$ and $V(\mathbf{x})$ for all $\mathbf{x} \in \mathcal{Q}_t$;
    \State $\mathcal{R}_t \gets \mathcal{P}_{t-1} \cup \mathcal{Q}_t$; 
    \State $(\mathcal{F}_1, \dots) \gets$ \Call{NonDominatedSorting}{$\mathcal{R}_t, \text{CDP}$};
    
    \State $\mathcal{P}_t \gets \emptyset, i \gets 1$;
    \While{$|\mathcal{P}_t| + |\mathcal{F}_i| \leq N_{\text{pop}}$}
        \State \Call{CrowdingDistance}{$\mathcal{F}_i$};
        \State $\mathcal{P}_t \gets \mathcal{P}_t \cup \mathcal{F}_i$;
        \State $i \gets i + 1$;
    \EndWhile
    \State \Call{CrowdingDistance}{$\mathcal{F}_i$};
    \State Sort $\mathcal{F}_i$ in descending order of crowding distance;
    \State $\mathcal{P}_t \gets \mathcal{P}_t \cup \mathcal{F}_i[1 : N_{\text{pop}} - |\mathcal{P}_t|]$;
\EndFor
\State \Return Non-dominated solutions in $\mathcal{F}_1$ of the final generation $\mathcal{P}_{N_{\text{gen}}}$.
\end{algorithmic}
\end{algorithm}

\subsubsection{Solution Selection}The proposed algorithm outputs a set of Pareto-optimal solutions $\mathcal{F}_1 = \{\mathbf{x}_1, \dots, \mathbf{x}_M\}$. {Each candidate solution $\mathbf{x}_i \in \mathcal{F}_1$ represents a different trade-off between the two conflicting objectives $f_1$ (channel gain) and $f_2$ (CS). To proceed with precoder design, a single operating point must be selected from this set. Therefore, the goal is to identify the solution that achieves the best overall balance between these two metrics.
Since $f_1$ and $f_2$ have different physical meanings and scales, they are first normalized to the range $[0, 1]$. In this context, a normalized value of zero represents the most desirable performance (i.e., maximum channel gain for $f_1$ and minimum CS for $f_2$), while a value of one represents the worst case within the Pareto set. The normalized objectives $\tilde{f}_m(\mathbf{x}_i)$ are defined as

\begin{minipage}{0.2\textwidth}
    \begin{equation}
    \hspace{-0.65cm}
        \tilde{f}_1(\mathbf{x}_i) = \frac{f_1^{\max} - f_1(\mathbf{x}_i)}{f_1^{\max} - f_1^{\min}}, \nonumber
    \end{equation}
    \vspace{0.02cm}
\end{minipage}
\hfill
\begin{minipage}{0.2\textwidth}
    \begin{equation}
        \tilde{f}_2(\mathbf{x}_i) = \frac{f_2(\mathbf{x}_i) - f_2^{\min}}{f_2^{\max} - f_2^{\min}}, \nonumber
    \end{equation}
    \vspace{0.02cm}
\end{minipage}

\noindent  where $f_m^{\max}$ and $f_m^{\min}$ denote the maximum and minimum values of the $m$-th objective over $\mathcal{F}_1$. Based on these normalized values, a joint quality score is defined to quantify the overall performance of each candidate solution. Specifically, $(1 - \tilde{f}_1(\mathbf{x}_i))$ and $(1 - \tilde{f}_2(\mathbf{x}_i))$ measure the closeness of $\mathbf{x}_i$ to the ideal performance for each objective. Their product thus reflects a balanced preference for solutions that simultaneously achieve high channel gain and low CS. The optimal operating point is then selected as
\begin{equation}
    \mathbf{x}^* = \arg \max_{\mathbf{x}_i \in \mathcal{F}_1} \left\{ (1 - \tilde{f}_1(\mathbf{x}_i)) \cdot (1 - \tilde{f}_2(\mathbf{x}_i)) \right\}.
\end{equation}
The resulting solution $\mathbf{x}^*$ is subsequently used as the final joint user–LED partition for precoder design.}

\subsection{Complexity Analysis}
The computational complexity of the proposed framework is evaluated in terms of the population size $N_{\text{pop}}$, the number of generations $N_{\text{gen}}$, and the per-generation computational cost. The dominant complexity in the fitness evaluation stage arises from the channel similarity (CS) computation, which involves pairwise channel correlations. For a single candidate solution, this operation incurs a complexity of $\mathcal{O}(K^2 N_T)$, resulting in an overall evaluation cost of $\mathcal{O}(N_{\text{pop}} K^2 N_T)$ per generation. In addition, population ranking is performed using the fast non-dominated sorting procedure, which has a worst-case complexity of $\mathcal{O}(M N_{\text{pop}}^2)$ \cite{Deb2002}, where $M=2$ denotes the number of objectives. The computational overhead of the remaining genetic operators, including selection, crossover, mutation, and repair, scales linearly with the population size and is therefore comparatively negligible. Aggregating these components over $N_{\text{gen}}$ generations, the total computational complexity of the proposed algorithm is given by
\begin{equation}
    \mathcal{T}_{\text{NSGA-II}} \approx \mathcal{O}\left( N_{\text{gen}} \left( N_{\text{pop}} K^2 N_T + N_{\text{pop}}^2 \right) \right).
\end{equation}
This confirms that the proposed framework operates in polynomial time, thereby avoiding the exponential complexity $\mathcal{O}(2^{K+N_T})$ associated with exhaustive search approaches.

\section{Simulation Results and Discussions}
\label{sec:simu}
This section presents simulation results to validate the performance of the two proposed design algorithms, as well as the effectiveness of the CS-reduction clustering method. The simulation considers a typical indoor environment with room dimensions of $5\text{m} \,\times 5\text{m} \,\times 3\text{m}$ (length $\times$ width $\times$ height), where all users are located on a horizontal plane at a height of $0.5\text{m}$. A three-dimensional Cartesian coordinate system is adopted, with its origin at the center of the room floor, while all LED luminaires are mounted on the ceiling. Unless otherwise specified, the simulation parameters are adopted from \cite{Thang2024}, with $P_t$ set to $30 \;\text{dBm}$ and the RS power allocation ratio set to a default value of $\rho = 2$.

% \begin{table}[ht]
% \caption{Simulation Parameters.}
% \label{tab:simparams}
% \begin{tabularx}{\columnwidth}{@{} X l @{}} 
% \toprule
% \multicolumn{2}{c}{\textbf{LED parameters}} \\
% \midrule
% % --- Transmitter ---
% LED bandwidth, $B$ & 20 MHz \\
% Semi-angle of half power, $\Phi_{1/2}$ & $60^\circ$ \\
% Electrical-to-optical conversion coefficient, $\eta$ & $2~\mathrm{W/A}$ \\
% Average emitted optical power per LED, $\overline{P}_n^s$ & $30~\mathrm{dBm}$ \\
% \midrule
% \multicolumn{2}{c}{\textbf{PD parameters}} \\
% \midrule
% % --- Receiver ---
% PD detect area, $A_r$ & $1~\mathrm{cm}^2$ \\
% FOV of the PD, $\Psi$ & $60^\circ$ \\
% PD responsivity, $\gamma$ & $0.54~\mathrm{A/W}$ \\
% Refractive index of concentrator, $\kappa$ & $1.5$ \\
% Optical filter gain, $T_s$ & $1$ \\
% Equivalent resistance, $\xi$ & $3~\Omega$ \\
% \midrule
% \multicolumn{2}{c}{\textbf{Other parameters}} \\
% \midrule
% % --- Channel/Environment ---
% Pre-amp noise density, $i_{\mathrm{amb}}$ & $5~\mathrm{pA}/\sqrt{\mathrm{Hz}}$ \\
% Ambient light photocurrent, $\chi_{\mathrm{amb}}$ & $10.93~\mathrm{A/(m^2 \cdot sr)}$ \\
% {Convergence thresholds, $\varepsilon_1, \varepsilon_2$} & $10^{-3}$ \\
% Private rate threshold, $\lambda_k$ & $0.2~\mathrm{bits/s/Hz}$ \\
% RS power allocation ratio, $\rho$ & $2$ \\
% Total power of private and common messages, $P_t$ & $30~\mathrm{dBm}$ \\
% \bottomrule
% \end{tabularx}
% \end{table}

\subsection{CCCP and CCCP-SDR Algorithms}
\begin{figure}[ht]
    \centering
    \begin{subfigure}[b]{0.32\linewidth} 
        \centering
        \includegraphics[width=\linewidth]{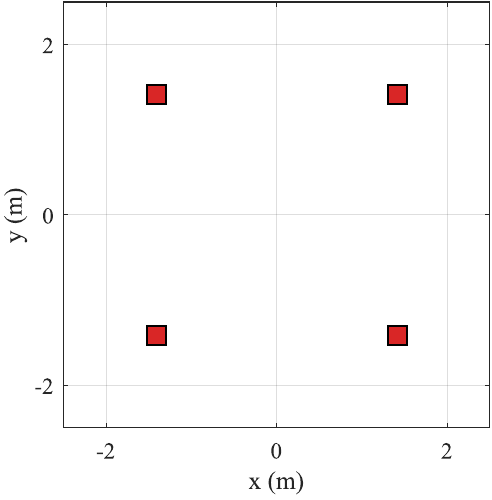}
        \caption{2$\times$2 layout.}
        \label{fig:layout_nt4}
    \end{subfigure}
    \hfill 
    \begin{subfigure}[b]{0.32\linewidth}
        \centering
        \includegraphics[width=\linewidth]{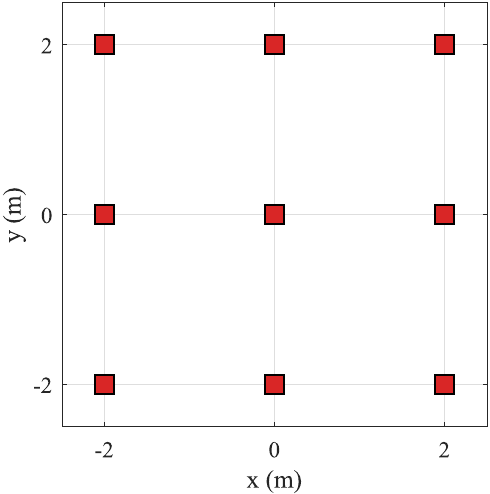}
        \caption{3$\times$3 layout.}
        \label{fig:layout_nt9}
    \end{subfigure}
    \hfill
    \begin{subfigure}[b]{0.32\linewidth}
        \centering
        \includegraphics[width=\linewidth]{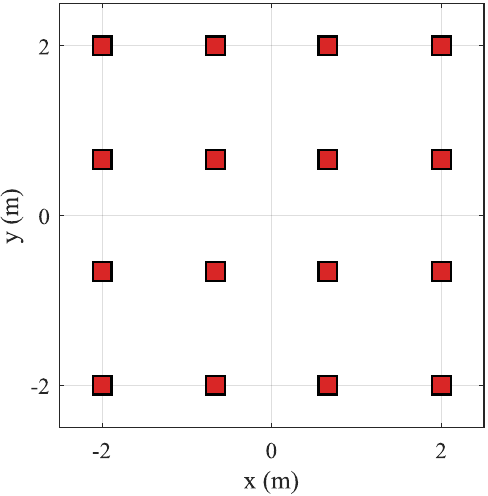}
        \caption{4$\times$4 layout.}
        \label{fig:layout_nt16}
    \end{subfigure} 
    \caption{Different layouts of LED transmitters.}
    \label{fig:led_layouts}
    \vspace{-\baselineskip}
\end{figure}

\begin{figure*}[!t]
    \centering
    \includegraphics[width=0.85\linewidth]{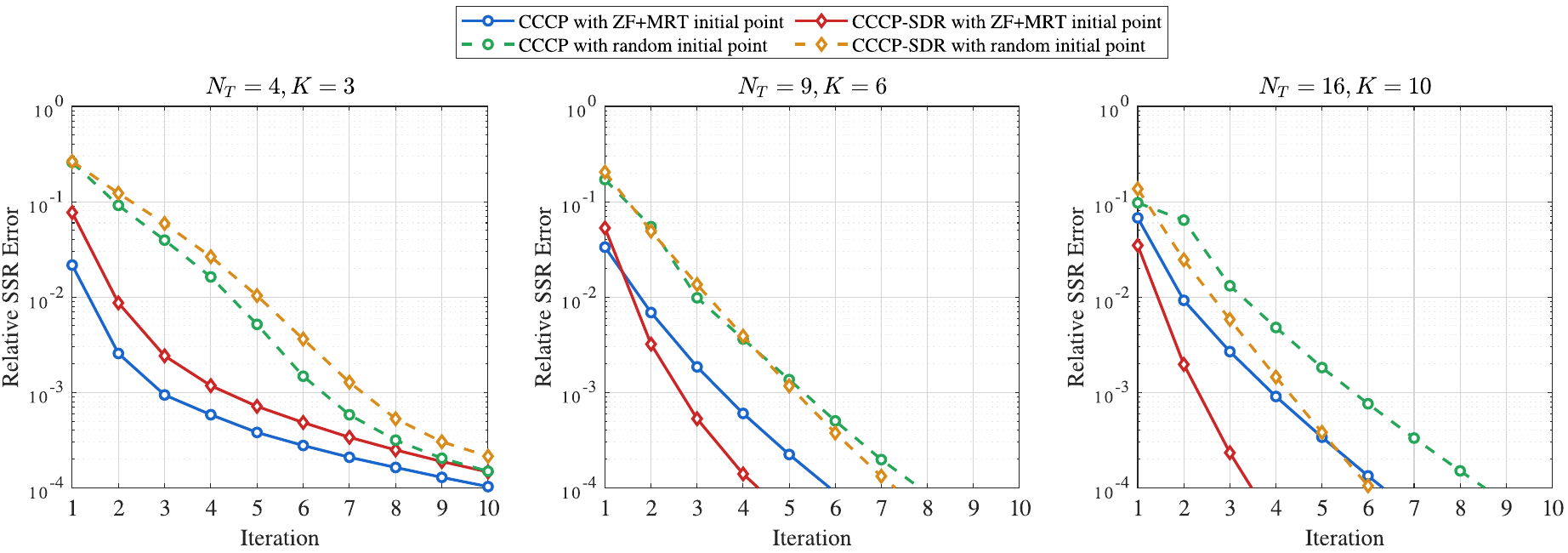}
    \caption{Convergence behaviors of proposed algorithms for different numbers of $N_T$ and $K$.}
    \label{fig:relative_error}
    \vspace{-\baselineskip}
\end{figure*}

First, the convergence behavior of the two proposed precoding schemes is evaluated in terms of the relative SSR error, as illustrated in Fig.~\ref{fig:relative_error}. The simulations consider three network configurations, $(N_T, K) \in \{(4,3), (9,6), (16,10)\}$, as shown in Fig.~\ref{fig:led_layouts}, with results averaged over 10,000 random channel realizations. 

For random precoder initialization, the CCCP approach exhibits slightly faster convergence than the CCCP-SDR method in the small-scale system ($N_T = 4$). However, this trend reverses as the system scales up to $N_T = 16$. Specifically, to reach a relative error of $10^{-3}$, both algorithms require approximately 7–8 iterations for $N_T = 4$, while for larger systems ($N_T = 9,~16$), \textbf{Algorithm 2} converges faster, requiring only around 6 iterations. It is also noted that random initialization leads to slower convergence and higher variance, which may introduce undesirable latency in practical implementations. To address this issue, we propose a heuristic initialization strategy that employs Zero-Forcing (ZF) for private streams and Maximum Ratio Transmission (MRT) beamforming for common messages. The results show a significant improvement in the convergence speed, where for $N_T = 4$, \textbf{Algorithms 1} and \textbf{2} reach the error of $10^{-3}$ approximately in 3 and 5 iterations, respectively. Furthermore, Table~\ref{table:execution_time} reveals a growing efficiency gap between the two schemes. While the execution time of the CCCP approach increases drastically with the system size, the CCCP-SDR maintains a manageable computational overhead, providing a significantly faster execution in the high-density configurations. This confirms that the SDR-based approach is not only more robust against network scaling but also better suited for latency-sensitive, real-time secure communications.

\begin{table}[ht]
\centering
\caption{\small{Average execution time (seconds) of each iteration.}}
\label{table:execution_time}
\begin{tabular}{|c|c|c|}
\hline
\diagbox{Setting}{Approach} & \begin{tabular}[c]{@{}c@{}}CCCP\\ \textbf{(Alg. 1)}\end{tabular} & \begin{tabular}[c]{@{}c@{}}CCCP-SDR\\ \textbf{(Alg. 2)}\end{tabular} \\ \hline
$N_T = 4$ \& $K = 3$ & 1.4668 & 0.7748 \\ \hline
$N_T = 9$ \& $K = 6$ & 3.3668 & 1.1993 \\ \hline
$N_T = 16$ \& $K = 10$ & 7.7534 & 2.2643 \\ \hline
\end{tabular}
\vspace{-\baselineskip}
\end{table}

\begin{figure}[ht]
    \centering
    \includegraphics[width=0.85\linewidth]{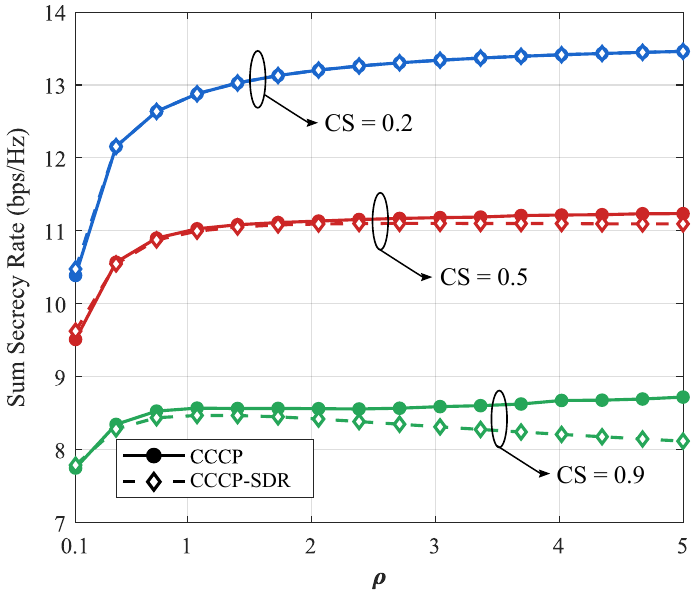}
    \caption{SSR versus $\rho$ for various CS levels.}
    \label{fig:SSR_rho}
    \vspace{-\baselineskip}
\end{figure}

Next, in Fig.~\ref{fig:SSR_rho}, the impact of the power ratio $\rho$ on the SSR is studied under different interference regimes (i.e., CS $= 0.2,~ 0.5$, and $0.9$). It is observed that the SSR increases with $\rho$, and gradually saturates beyond $\rho \approx 2$. Meanwhile, the maximum SSR decreases from $13.4$ to $8.7$ bps/Hz as the CS increases from $0.2$ to $0.9$, reflecting the adverse effect of stronger spatial correlation. 
In addition, for moderate interference levels CS $\le 0.5$, \textbf{Algorithm 1} and \textbf{Algorithm 2} achieve nearly identical performance. However, under strong interference $CS = 0.9$, a noticeable performance gap emerges and widens with increasing $\rho$, reaching approximately $0.6$ bps/Hz at $\rho = 5$. This divergence can be attributed to the broader feasibility region of \textbf{Algorithm 2}. Specifically, by successfully identifying feasible precoding solutions across a wider range of channel realizations—including those where \textbf{Algorithm 1} fails to converge—the ensemble average of \textbf{Algorithm 2} incorporates more low-rate instances, which in turn slightly reduces its mean SSR.

\begin{figure}[t]
    \centering
    \includegraphics[width=0.85\linewidth]{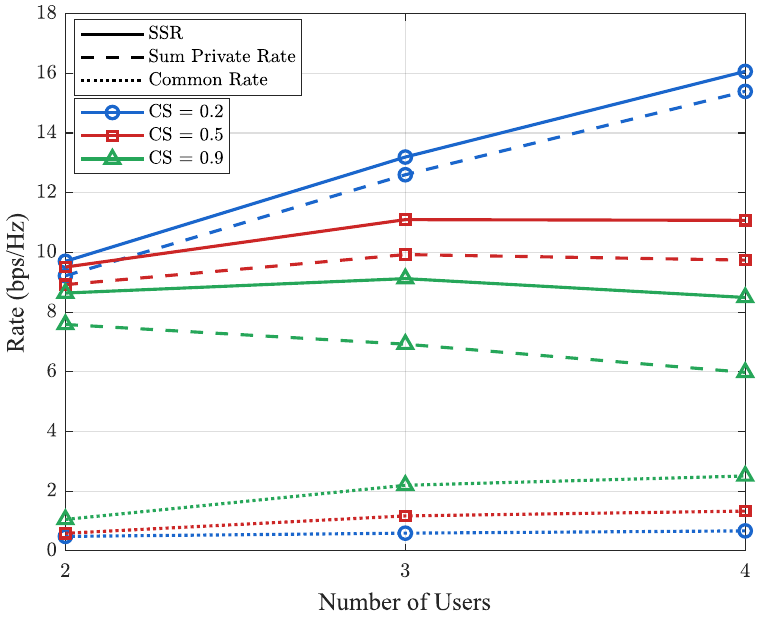}
    \caption{SSR and its rate components versus the number of users under different CS levels.}
    \label{fig:SSR_CS}
    \vspace{-\baselineskip}
\end{figure}

Building upon the impact of power allocation discussed above, Fig.~\ref{fig:SSR_CS} further examines the scalability of the system as the number of users $K$ increases from 2 to 4. Although the SSR generally improves with $K$ due to multi-user diversity, the achievable gains remain strongly dependent on the level of spatial correlation. For $CS = 0.2$, the SSR increases significantly from $10$ to approximately $16$ bps/Hz, driven primarily by private streams, while the common rate remains negligible (below $1$ bps/Hz). In contrast, under high-interference conditions ($CS = 0.9$), the SSR saturates at around $8.5$ bps/Hz, as the sum private rate decreases from $7.5$ to $6$ bps/Hz due to the limited spatial DoF. Consistent with the observations in Fig.~\ref{fig:SSR_rho}, RSMA compensates for this limitation by re-allocating power to the common message, increasing its contribution from approximately $1$ to $2.5$ bps/Hz at $K=4$. These results further confirm the effectiveness of RSMA in dense VLC networks, where it sustains reliable secrecy performance by adaptively exploiting the power domain when spatial DoF becomes constrained.

\begin{figure}[ht]
    \centering
    \includegraphics[width=0.85\linewidth]{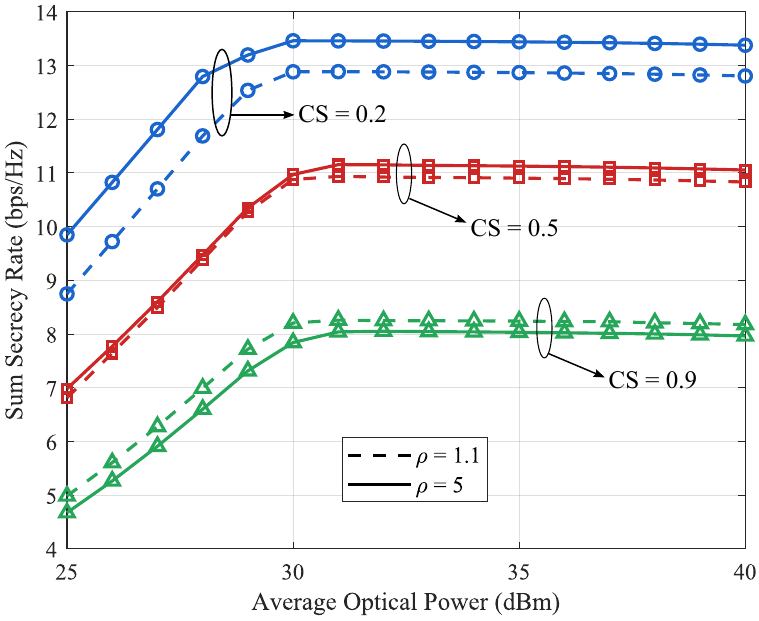}
    \caption{SSR versus Average Optical Power and CS levels.}
    \label{fig:SSR_Ps}
    \vspace{-\baselineskip}
\end{figure}

Extending the previous analysis, Fig.~\ref{fig:SSR_Ps} shows the SSR versus the transmit power of each LED luminaire for different CS values, with $\rho$ selected as $1.1$ and $5$. The SSR increases approximately linearly with transmit power before reaching a saturation point, occurring around $30$ dBm for CS $=0.2$ and shifting to about $32$ dBm under higher correlation conditions. Consistent with the trends observed earlier, the impact of $\rho$ depends strongly on the spatial correlation. In low-correlation regimes, $\rho=1.1$ provides better performance, whereas under higher correlation (CS = $0.5$ and $0.9$), $\rho=5$ becomes more effective. This shift reflects the changing balance between common and private message contributions as interference increases. Taken together, these findings underscore the critical interplay between transmit power, power allocation, and spatial correlation, highlighting the importance of adaptive RSMA strategies for maintaining robust secrecy performance across diverse channel conditions.

\begin{figure}[ht]
    \centering
    \includegraphics[width=0.85\linewidth]{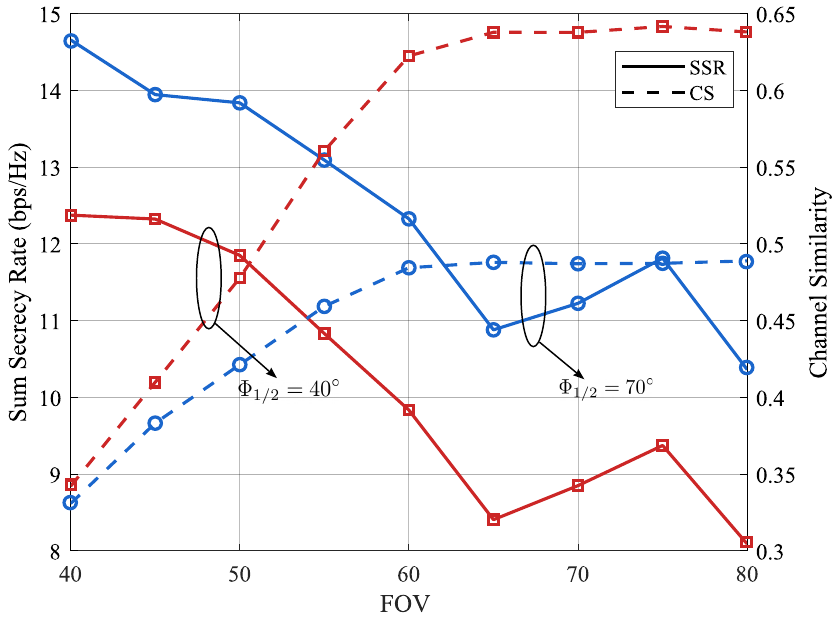}
    \caption{SSR and CS versus receiver FOV for different $\Phi_{1/2}$.}
    \label{fig:SSR_FOV_SA}
    \vspace{-\baselineskip}
\end{figure}

In addition, the impact of receiver's FOV on system secrecy performance is studied in Fig.~\ref{fig:SSR_FOV_SA}, for given different values of $\Phi_{1/2}$. The SSR decreases with increasing FOV, dropping from $14.7$ to $10.4$ bps/Hz for $\Phi_{1/2} = 40^\circ$, and from $12.4$ to $8.1$ bps/Hz for $\Phi_{1/2} = 70^\circ$. {This behavior can be explained by the fact that a wider FOV allows receivers to capture signals from a larger and more overlapping set of LEDs, increasing channel similarity among users. As a result, the transmitter’s ability to spatially separate intended signals and suppress information leakage is reduced, leading to a degradation in SSR.} Moreover, a notable decoupling effect is observed beyond an FOV of $60^\circ$, that is while the CS tends to saturate, the SSR continues to decline. This indicates that, at wide FOVs, secrecy degradation is primarily driven by the reduction in Lambertian optical power density rather than further increases in spatial correlation. Besides, the narrower beam ($\Phi_{1/2} = 40^\circ$) consistently achieves higher secrecy performance compared to the wider configuration. The localized fluctuations observed around $65^\circ$ and $75^\circ$ can be attributed to the geometric interaction between the 4-LED array and the photodetector’s angular response.

\subsection{CS-Reduction Clustering Design}
In this sub-section, we investigate the performance of the proposed joint clustering framework using NSGA-II. To establish a rigorous baseline, we adopt the clustering strategies in \cite{Zhang2016, Li2017, Yang2022}-hereafter referred to as the Conventional UC-Clustering (C-UCC) and integrate them with our secrecy-aware design. Although these schemes were originally developed based on distance-based associations, this integration enables a fair comparison of how different cluster strategies impact the overall SSR.

\begin{figure}[ht]
    \centering
    \includegraphics[width=0.85\linewidth]{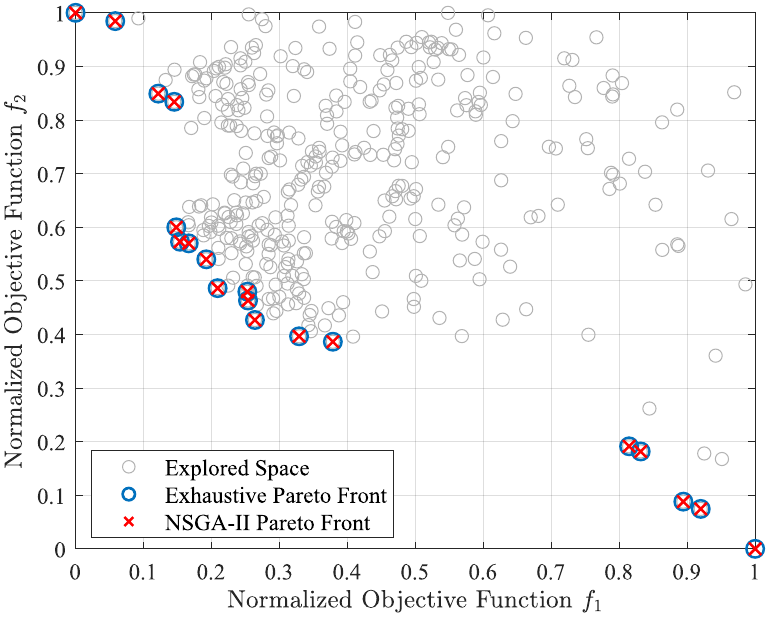}
    \caption{Pareto front of NSGA-II and exhaustive search.}
    \label{fig:pareto_front}
    \vspace{-\baselineskip}
\end{figure}
First, the effectiveness of the proposed NSGA-II as a high-fidelity candidate selection mechanism is demonstrated in Fig.~\ref{fig:pareto_front}, where its Pareto front closely matches that obtained by exhaustive search for the normalized objectives $f_1$ and $f_2$. By accurately capturing the boundary defined by the exhaustive method, NSGA-II efficiently reduces the large search space to a manageable set of non-dominated solutions. Notably, the resulting Pareto front does not guarantee strict optimality in terms of achievable rate. Rather, it identifies high-potential clustering configurations that align with the defined objectives, enabling the system to bypass suboptimal solutions and focus on the most relevant regions of the optimization landscape.

\begin{figure}[ht]
    \centering
    \includegraphics[width=0.85\linewidth]{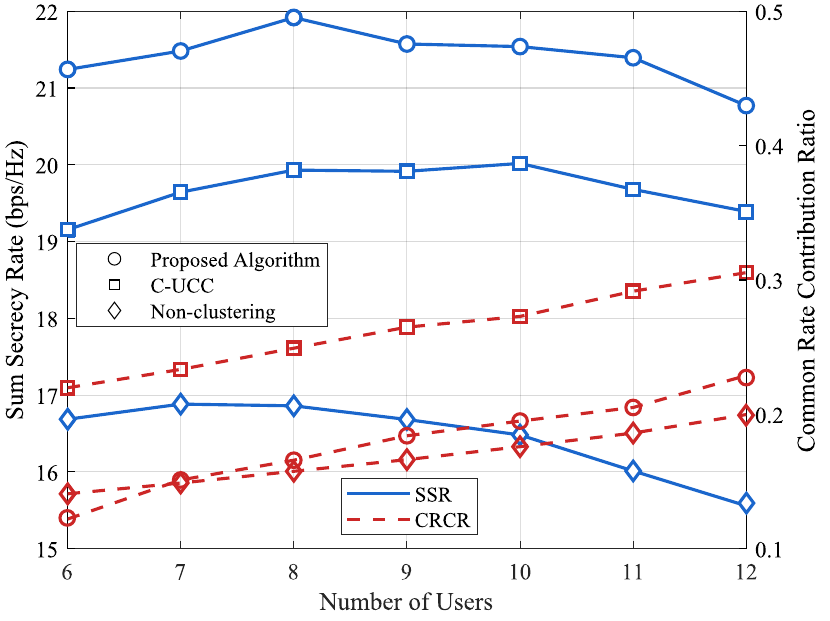}
    \caption{SSR and Common Rate Contribution Ratio versus the number of users for different schemes.}
    \label{fig:SSR_SC_NSGA2_numUser}
    \vspace{-\baselineskip}
\end{figure}
Building on the effectiveness of NSGA-II in identifying high-quality clustering candidates, Fig.~\ref{fig:SSR_SC_NSGA2_numUser} evaluates the performance of the proposed algorithm and compares with that of C-UCC and non-clustering strategy as the number of users $K$ increases from 6 to 12. It is observed that both the proposed and C-UCC schemes initially benefit from multi-user diversity, with SSR peaking at $K=8$, followed by a gradual decline as inter-cluster interference limits the available spatial DoF. Nevertheless, the proposed method consistently outperforms the benchmark by about $2$ bps/Hz across this range, while the non-clustering baseline not only degrades sharply beyond $K=7$ but also yields the lowest performance overall, due to its inability to manage inter-user interference without structured clustering.
Furthermore, the Common Rate Contribution Ratio (CRCR), i.e., $R_c/\text{SSR}$, reveals that the proposed scheme achieves higher secrecy rates despite relying less on common signaling. This highlights the effectiveness of CS-aware clustering in preserving spatial separability of private streams, thereby reducing overhead and sustaining secrecy performance in dense regimes.

\begin{figure}
    \centering
    \includegraphics[width=0.85\linewidth]{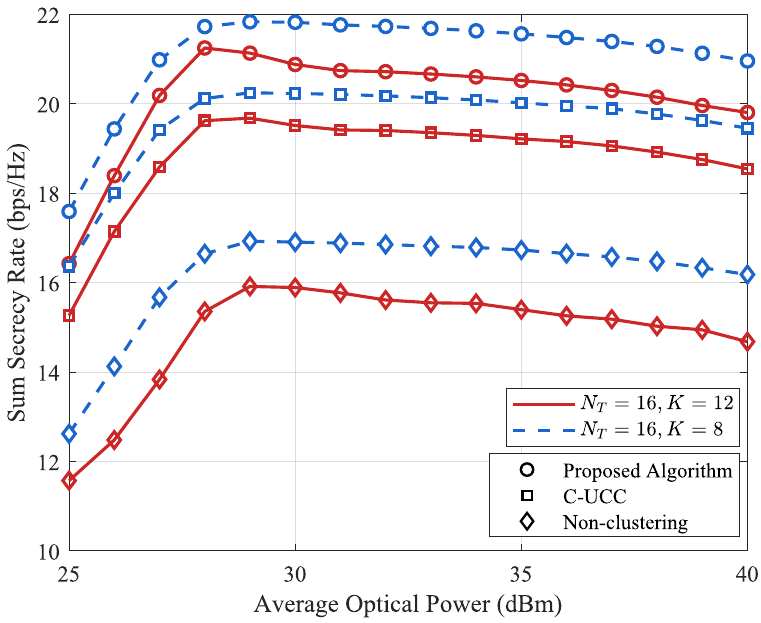}
    \caption{SSR versus Average Optical Power for different schemes.}
    \label{fig:SSR_SC_NSGA2_power}
    \vspace{-\baselineskip}
\end{figure}

Finally, in Fig.~\ref{fig:SSR_SC_NSGA2_power}, the SSR performance of the two algorithms and the non-clustering scheme is compared in terms of average optical power. In all cases, the SSR increases with transmit power and reaches a peak around 28--30 dBm, beyond which a slight degradation is observed. This behavior indicates an optimal operating region, where further power increases mainly amplify interference rather than improving the useful signal strength. Across both user settings, $K=8$ and $K=12$, the proposed algorithm again achieves the highest SSR, followed by the C-UCC scheme, while the non-clustering approach performs worst throughout. In addition, the lower SSR for $K=12$ compared to $K=8$ confirms that spatial correlation remains the dominant bottleneck for SSR.

\section{Conclusion}
In this paper, we investigated the SSR performance of RSMA-based VLC systems, specifically accounting for the detrimental impact of spatial correlation. To address the non-convex nature of the optimization problem, two sub-optimal solutions based on CCCP and SDR were developed for precoder design. Furthermore, to mitigate severe inter-user interference and information leakage in dense deployments, a novel CSR clustering strategy optimized via the NSGA-II was proposed. Numerical results obtained across various network configurations confirm that high CS scenarios lead to a significant degradation of the SSR. They also confirm that the proposed CS-aware clustering approach achieves a notable SSR gain of approximately 2 bps/Hz compared to conventional distance-based clustering methods. 
\label{sec:conclusion}

\bibliographystyle{IEEEtran}
\bibliography{references}

\end{document}